\def\year{2023}\relax
\documentclass[letterpaper]{article} 
\usepackage{aaai24}

\def\B{{\bf B}}

\def\E{{\mathbb{E}}}

\def\L{{\bf L}}

\def\S{{\bf S}}

\def\T{{\bf T}}

\def\0{{\bf 0}}
\def\1{{\bf 1}}

\def\OM{{\mathcal O}}

\def\Ome{\mbox{\boldmath$\Omega$\unboldmath}}

\def\argmin{\mathop{\rm argmin}}

\def\Lsize{\hbox{\space \raise-2mm\hbox{$\textstyle \L \atop \scriptstyle {m\times 3n}$} \space}}
\def\Ssize{\hbox{\space \raise-2mm\hbox{$\textstyle \S \atop \scriptstyle {m\times 3n}$} \space}}
\def\Osize{\hbox{\space \raise-2mm\hbox{$\textstyle \Ome \atop \scriptstyle {m\times 3n}$} \space}}
\def\Tsize{\hbox{\space \raise-2mm\hbox{$\textstyle \T \atop \scriptstyle {3n\times n}$} \space}}
\def\Bsize{\hbox{\space \raise-2mm\hbox{$\textstyle \B \atop \scriptstyle {m\times n}$} \space}}

\newcommand{\ip}[2]{\left\langle #1, #2 \right \rangle}

\newcommand{\TheName}{FedASMU}

\usepackage{helvet}  
\usepackage{courier}  
\usepackage[hyphens]{url}  
\usepackage{graphicx} 
\usepackage{natbib}  
\usepackage{caption} 
\DeclareCaptionStyle{ruled}{labelfont=normalfont,labelsep=colon,strut=off} 
\usepackage{algorithm}
\usepackage{algorithmic}
\usepackage{amsmath} 
\usepackage{amsfonts} 
\usepackage{amssymb} 
\usepackage{amsthm}
\usepackage{mathrsfs}
\newtheorem{theorem}{Theorem}

\newtheorem{assumption}{Assumption}

\usepackage{xcolor}
\usepackage{subfigure}
\usepackage{booktabs}
\usepackage{multirow}
\usepackage{multicol}
\usepackage{color}
\usepackage{textpos}
\usepackage{newfloat}
\usepackage{listings}
\floatstyle{ruled}
\newfloat{listing}{tb}{lst}{}
\floatname{listing}{Listing}

\title{\TheName{}: Efficient Asynchronous Federated Learning with Dynamic Staleness-aware Model Update}

\author {
    Ji Liu$^{1, }$\footnote{Corresponding author (jiliuwork@gmail.com and jiajuncheng@suda.edu.cn).},
    Juncheng Jia$^{2, *}$,
    Tianshi Che$^3$,
    Chao Huo$^2$,
    Jiaxiang Ren$^3$,
    Yang Zhou$^3$,
    Huaiyu Dai$^4$,
    Dejing Dou$^5$
}
\affiliations {
    $^1$ Hithink RoyalFlush Information Network Co., Ltd., $^2$ Soochow University, $^5$ Boston Consulting Group, China, \\
    $^3$ Auburn University, $^4$ North Carolina State University, United States.
}

\begin{document}

\maketitle

\begin{textblock*}{10cm}(5.5cm,18cm) 
   {\huge To appear in AAAI 2024}
\end{textblock*}

\begin{abstract}

As a promising approach to deal with distributed data, Federated Learning (FL) achieves major advancements in recent years. FL enables collaborative model training by exploiting the raw data dispersed in multiple edge devices. However, the data is generally non-independent and identically distributed, i.e., statistical heterogeneity, and the edge devices significantly differ in terms of both computation and communication capacity, i.e., system heterogeneity. The statistical heterogeneity leads to severe accuracy degradation while the system heterogeneity significantly prolongs the training process. In order to address the heterogeneity issue, we propose an Asynchronous Staleness-aware Model Update FL framework, i.e., \TheName{}, with two novel methods. First, we propose an asynchronous FL system model with a dynamical model aggregation method between updated local models and the global model on the server for superior accuracy and high efficiency. Then, we propose an adaptive local model adjustment method by aggregating the fresh global model with local models on devices to further improve the accuracy. Extensive experimentation with 6 models and 5 public datasets demonstrates that \TheName{} significantly outperforms baseline approaches in terms of accuracy (0.60\% to 23.90\% higher) and efficiency (3.54\% to 97.98\% faster).

\end{abstract}

\section{Introduction}

In recent years, numerous edge devices have been generating large amounts of distributed data.
Due to the implementation of laws and regulations, e.g., General Data Protection Regulation (GDPR) \cite{GDPR},
the traditional training approach, which aggregates the distributed data into a central server or a data center, becomes almost impossible. As a promising approach, Federated Learning (FL) \cite{mcmahan2021advances,liu2022distributed}
enables collaborative model training by transferring gradients or models instead of raw data. FL avoids privacy or security issues incurred by direct raw data transfer while exploiting multiple edge devices to train a global model. FL has been applied in diverse areas, such as computer vision \cite{liu2020fedvision}, nature language processing \cite{liu2021federated}, bioinformatics \cite{chen2021fl}, 
and healthcare \cite{nguyen2022federated}.

Traditional FL typically exploits a parameter server (server) \cite{li2014scaling,liu2021heterps} to coordinate the training process on each device with a synchronous \cite{mcmahan2017communication,Li2020FedProx,liu2023distributed,Jia2023Efficient}  mechanism. The synchronous training process generally consists of multiple rounds and each round contains five steps. First, the server selects a set of devices \cite{shi2020joint}. Second, the server broadcasts the global model to the selected devices. Third, local training is carried out with the data in each selected device. Fourth, each device uploads the updated model (gradients) to the server. Fifth, the server aggregates the uploaded models to generate a new global model when all the selected devices complete the aforementioned four steps. Although the synchronous mechanism is effective and simple to implement, stragglers may significantly prolong the training process \cite{jiang2022pisces} with heterogeneous devices \cite{lai2021oort,yang2021characterizing}. Powerful devices may remain idle when the server is waiting for stragglers \cite{wu2020safa}, incurring significant efficiency degradation.

Within the FL paradigm, the devices are typically highly heterogeneous in terms of computation and communication capacity \cite{li2018learning,wu2020safa,nishio2019client,che2022federated,che2023fast} and data distribution \cite{mcmahan2017communication,Li2020FedProx,Wang2020Tackling,che2023federated}. Some devices may complete the local training and update the model within a short time, while some other devices may take a much longer time to finish this process and may fail to upload the model because of modest bandwidth or high latency, which is denoted by system heterogeneity. In addition, the data in each device may be non-Independent and Identically Distributed (non-IID) data, which refers to statistical heterogeneity. The statistical heterogeneity can lead to diverse local objectives \cite{Wang2020Tackling} and client drift \cite{Karimireddy2020SCAFFOLD,hsu2019measuring} issues, which degrades the accuracy of the global model in FL. 

Asynchronous FL \cite{xu2021asynchronous,wu2020safa,nguyen2022federated} enables the server to aggregate the uploaded models without waiting for stragglers, which improves the efficiency. However, this mechanism may encounter low accuracy brought by stale uploaded models and non-IID data \cite{zhou2021tea}. For instance, when a device uploads a model updated based on an old global model, the global model has already been updated multiple times. Then, the simple aggregation of the uploaded model may drag the global model to previous status, which corresponds to inferior accuracy. In addition, the asynchronous FL mechanism may fail to converge \cite{Su2022How} due to the lack of the staleness control \cite{xie2019asynchronous}. 

Existing works 
address the system heterogeneity and the statistical heterogeneity separately. Some works focus on device scheduling \cite{shi2020joint,shi2020device,wu2020safa,zhou2022efficient,liu2022multi} to avoid the inefficiency incurred by stragglers while this mechanism may correspond to inferior accuracy due to insufficient participation of devices. Asynchronous FL approaches are proposed to mitigate the system heterogeneity while they either exploit static polynomial formula to deal with the staleness \cite{xie2019asynchronous,Su2022How,chen2021fedsa} or leverage simple attention mechanism \cite{chen2020asynchronous}. However, they cannot dynamically adjust the importance of each uploaded model within the model aggregation process, which leads to modest accuracy. Some other approaches introduce regularization \cite{Li2020FedProx}, gradient normalization \cite{Wang2020Tackling}, and momentum methods \cite{hsu2019measuring,JinAccelerated2022} to address the statistical heterogeneity, while they focus on synchronous FL. 

In this paper, we propose an original Asynchronous Federated learning framework with Staleness-aware Model Update (\TheName{}). To address the system heterogeneity, we design an asynchronous FL system and propose a dynamical adjustment method to update the importance of updated local models and the global model based on both the staleness and the local loss for superior accuracy and high efficiency. We enable devices to adaptively aggregate fresh global models so as to reduce the staleness of the local model. We summarize the major contributions in this paper as follows:

\begin{itemize}
    \item We propose a novel asynchronous FL system model with a dynamic model aggregation method on the server, which adjusts the importance of updated local models and the global model based on the staleness and the impact of local loss for superior accuracy and high efficiency.
    \item We propose an adaptive local model adjustment method on devices to integrate fresh global models into the local model so as to reduce staleness for superb accuracy. The model adjustment consists of a Reinforcement Learning (RL) method to select a proper time slot to retrieve global models and a dynamic method to adjust the local model aggregation.
    \item We conduct extensive experiments with 9 state-of-the-art baseline approaches, 6 typical models, and 5 public real-life datasets, which reveals \TheName{} can well address the heterogeneity issues and significantly outperforms the baseline approaches.
\end{itemize}

The rest of the paper is organized as follows. We present the related work in Section \ref{sec:relatedWork}. Then, we formulate the problem and explain the system model in Section \ref{sec:systemModel}. We propose the staleness-aware model update in Section \ref{sec:modelAggregation}. The experimental results are given in Section \ref{sec:exp}. Finally, Section \ref{sec:con} concludes the paper.

\section{Related Work}
\label{sec:relatedWork}

A bunch of FL approaches \cite{mcmahan2017communication,Li2020FedProx,Wang2020Tackling,Karimireddy2020SCAFFOLD,acar2021federated} have been designed to collaboratively train a global model utilizing the distributed data in mobile devices. Most of them \cite{bonawitz2019towards} exploit a synchronous mechanism to perform the model aggregation on the server. With the synchronous mechanism, the server needs to wait for all the selected devices to upload models before the model aggregation, which is inefficient because of stragglers. Due to diverse device availability and system heterogeneity, the probability of the occurrence of the straggler effect increases when the scale of devices becomes large \cite{li2014scaling}. 

Three types of approaches are proposed to address the system heterogeneity with the synchronous mechanism. The first type is to schedule proper devices to perform the local training process \cite{shi2020joint,shi2020device,wu2020safa} while considering the computation and communication capacity to achieve load balance among multiple devices. However, this approach may significantly reduce the participation frequency of some modest devices, which degrades accuracy. Second, pruning \cite{Zhang2022FedDUAP} or dropout \cite{horvath2021fjord} techniques are leveraged during the training process, while incurring lossy compression and modest accuracy. Third, the device clustering approach \cite{Li2022FedHiSyn} groups the devices of similar capacity into the same cluster, and utilizes a hierarchical architecture \cite{abad2020hierarchical} to perform model aggregation. All these approaches focus on the synchronous mechanism with low efficiency and may incur severe accuracy degradation due to statistical heterogeneity.

Multiple model aggregation methods \cite{Karimireddy2020SCAFFOLD,Mitra2021FedLin,sattler2020clustered} exist to handle the statistical heterogeneity with the synchronous mechanism. In particular, regularization \cite{Li2020FedProx,acar2021federated}, gradient normalization \cite{Wang2020Tackling}, classifier calibration \cite{Luo2021No}, and momentum-based \cite{hsu2019measuring,reddi2021adaptive} methods adjust the local objectives to reduce the accuracy degradation brought by heterogeneous data. Contrastive learning \cite{li2021model}, personalization \cite{Sun2021PartialFed,ozkara2021quped}, meta-learning-based method \cite{Khodak2019Adaptive}, and multi-task learning \cite{Smith2017Federated} adapt the global model or local models to non-IID data. However, these methods cannot dynamically adjust the importance of diverse models and only focus on the synchronous mechanism.

\begin{figure*}[!htbp]
\centering
\includegraphics[width=0.65\linewidth]{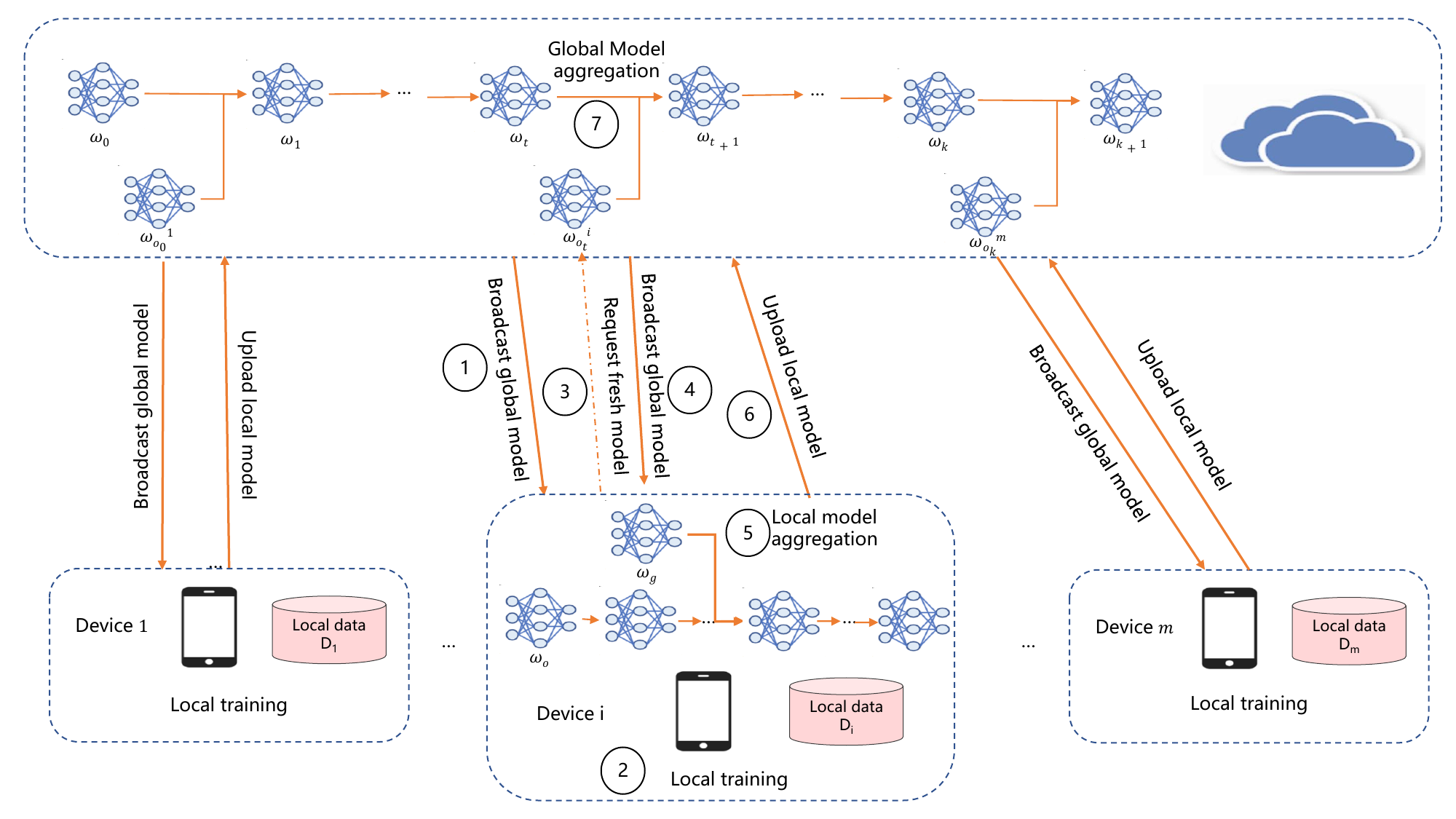}
\vspace{-4mm}
\caption{The system model of \TheName{}.}
\label{fig:system}
\vspace{-8mm}
\end{figure*}

To conquer the system heterogeneity problem, asynchronous FL \cite{xu2021asynchronous,nguyen2022federated} enables the global model aggregation without waiting for all the devices. The asynchronous FL can be performed once a model is uploaded from an arbitrary device \cite{xie2019asynchronous} or when multiple models are buffered \cite{Nguyen2022FedBuff}. However, the old uploaded models may drag the global model to a previous status, which significantly degrades the accuracy \cite{Su2022How}. Several methods are proposed to improve the accuracy of asynchronous FL. For instance, the impact of the staleness and the divergence of model updates is considered to adjust the importance of uploaded models \cite{Su2022How}, which cannot dynamically adapt the weights based on the training status, e.g., loss values. The attention mechanism and the average local training time are exploited to adjust the weights of uploaded models \cite{chen2020asynchronous} without the consideration of staleness. In addition, the uploaded model with severe staleness can be replaced by the latest global model \cite{wu2020safa} to reduce the impact of staleness while losing important information from the device. Furthermore, a staleness-based polynomial formula can be utilized to assign high weights to fresh models \cite{park2021sageflow,xie2019asynchronous,chen2021fedsa} while the loss value of the model can be leveraged to adjust the importance of models \cite{park2021sageflow}. 
However, these methods only consider static formulas, which cannot dynamically adjust the importance of models for the objective of minimizing the loss so as to improve the accuracy.

Different from the existing works, we propose an asynchronous FL framework, i.e., \TheName{}, to address the system heterogeneity. \TheName{} adjusts the importance of uploaded models based on the staleness while enabling devices to adaptively aggregate fresh global models to further mitigate the staleness issues, which handles the statistical heterogeneity. 

\section{Aysnchronous System Architecture}
\label{sec:systemModel}

In this section, we present the problem formulation for FL and the asynchronous system model.

We consider an FL setting composed of a powerful server and $m$ devices, denoted by $\mathcal{M}$, which collaboratively train a global model (the main notations are summarized in Appendix). Each device $i$ stores a local dataset $\mathcal{D}_i = \{\boldsymbol{x}_{i,d} \in \mathbb{R}^s, y_{i,d} \in \mathbb{R} \}_{d=1}^{D_i}$ with $D_i = |\mathcal{D}_i|$ data samples where $\boldsymbol{x}_{i,d}$ is the $d$-th $s$-dimensional input data vector, and $y_{i,d}$ is the label of $\boldsymbol{x}_{i,d}$. The whole dataset is denoted by $\mathcal{D} = \bigcup_{i \in \mathcal{M}} \mathcal{D}_i$ with $D=\sum_{i \in \mathcal{M}} D_i$. Then, the objective of the training process within FL is:
\vspace{-3mm}
\begin{equation}\label{eq:FLProblem}
\vspace{-3mm}
\min_{\boldsymbol{w}}\Big\{\mathcal{F}(\boldsymbol{w}) \triangleq \frac{1}{|D|} \sum_{i \in \mathcal{M}}|\mathcal{D}_i| \mathcal{F}_i(\boldsymbol{w}) \Big\}, \tag{$\mathcal{P}$}
\end{equation}
where $\boldsymbol{w}$ represents the global model, $\mathcal{F}_i(\boldsymbol{w})$ is the local loss function defined as $\mathcal{F}_i(\boldsymbol{w}) \triangleq \frac{1}{|D_i|}\sum_{\{\boldsymbol{x}_{i,d}, y_{i,d}\} \in \mathcal{D}_i}F(\boldsymbol{w}, \boldsymbol{x}_{i,d}, y_{i,d})$, and $F(\boldsymbol{w}, \boldsymbol{x}_{i,d}, y_{i,d})$ is the loss function to measure the error of the model parameter $\boldsymbol{w}$ on data sample $\{\boldsymbol{x}_{i,d}, y_{i,d}\}$. 

In order to address the problem defined in Formula \ref{eq:FLProblem}, we propose an asynchronous FL framework as shown in Figure \ref{fig:system}. The server triggers the local training of $m'$ devices with a constant time period $\mathcal{T}$. The training process is composed of multiple global rounds. At the beginning of the training, the version of the global model is 0. Then, after each global round, the version of the global model increases by 1. Each global round is composed of 7 steps. First, the server triggers $m' (m' \le m) $ devices and broadcasts the global model $\boldsymbol{w}_o$ to each device at Step \textcircled{1}. The $m'$ devices are randomly selected available devices. Then, each device performs local training with its local dataset at Step \textcircled{2}. During the local training process, Device $i$  requests a fresh global model (Step \textcircled{3}) from the server to reduce the staleness of the local training as the global model may be updated at the same time. Then, the server sends the global model $\boldsymbol{w}_g$ to the device at Step \textcircled{4}, if $\boldsymbol{w}_g$ is newer than $\boldsymbol{w}_o$, i.e., $g > o$. After receiving the fresh global model, the device aggregates the global model and the latest local model to a new model at Step \textcircled{5} and continues the local training with the new model. When the local training is completed, Device $i$ uploads the local model to the server at Step \textcircled{6}. Finally, the server aggregates the latest global model $\boldsymbol{w}_t$ with the uploaded model $\boldsymbol{w}^i_o$ at Step \textcircled{7}. When aggregating the global model $\boldsymbol{w}_t$ and the uploaded local model $\boldsymbol{w}^i_o$, the staleness of the local model is calculated as $\tau_i = t - o + 1$. When the staleness $\tau_i$ is significant, the local model may drag the global model to a previous version corresponding to inferior accuracy due to legacy information. We discard the uploaded local models when the staleness exceeds a predefined threshold $\tau$ to meet the staleness bound so as to ensure the convergence \cite{ho2013more}. 

\section{Staleness-aware Model Update}
\label{sec:modelAggregation}

In this section, we propose our dynamic staleness-aware model aggregation method on the server (Step \textcircled{7}) and the adaptive local model adjustment method on devices (Steps \textcircled{3} and \textcircled{5}). 

\subsection{Dynamic Model Update on the Server}
\label{subsec:MAServer}

In this subsection, we propose our dynamic staleness-aware model update method on the server. When the server receives an uploaded model $\boldsymbol{w}_o^i$ from Device $i$ with the original version $o$, it updates the current global model $\boldsymbol{w}_t$ according to the following formula: 
\vspace{-3mm}
\begin{equation}
\label{eq:serverAggregation}
\boldsymbol{w}_{t+1} = (1-\alpha^i_t) \boldsymbol{w}_{t} + \alpha^i_t \boldsymbol{w}_o^i,
\vspace{-1mm}
\end{equation}
where $\alpha^i_t$ represents the importance of the uploaded model from Device $i$ at global round $t$, which may have a significant impact on the accuracy of the aggregated model \cite{xie2019asynchronous}. Then, we decompose the problem defined in Formula \ref{eq:FLProblem} to the following bi-level optimization problem \cite{Bard1998PracticalBO}:
\vspace{-1mm}
\begin{align}\label{eq:biProblem}
\min_{\boldsymbol{w},\boldsymbol{A}}\Big\{&\mathcal{F}(\boldsymbol{w}, \boldsymbol{A}) \triangleq \frac{1}{|D|} \sum_{i \in \mathcal{M}}|\mathcal{D}_i| \mathcal{F}_i(\boldsymbol{w}(\boldsymbol{A})) \Big\}, \nonumber\\
&\boldsymbol{w}(\boldsymbol{A}) = (1-\alpha^i_t) \boldsymbol{w}_{t} + \alpha^i_t \boldsymbol{w}_o^i, \alpha^i_t \in \boldsymbol{A},
\tag{$\mathcal{P}1$}\\
\textit{s.t.} &~\boldsymbol{w}^i_o = \argmin_{\boldsymbol{w}^i_o}\mathcal{F}_i(\boldsymbol{w}^i_o)~\forall~i \in \mathcal{M}, \label{eq:Problem2} \tag{$\mathcal{P}2$}\\ 
&~\boldsymbol{A} = \argmin_{\boldsymbol{A}}\mathcal{F}(\boldsymbol{w}, \boldsymbol{A}), \label{eq:Problem3} \tag{$\mathcal{P}3$}
\vspace{-4mm}
\end{align}
where $\boldsymbol{A} = \{\alpha_t^1, \alpha_t^2, ..., \alpha_t^m\}$ is a set of values corresponding to the importance of the uploaded models from devices. Problem \ref{eq:Problem2} is the minimization of the local loss function, which is detailed in Section \ref{subsec:MADevice}. Inspired by \cite{xie2019asynchronous}, we propose a dynamic polynomial function to represent $\alpha^i_t$ defined in Formula \ref{eq:alpha} for Problem \ref{eq:Problem3}.
\vspace{-3mm}
\begin{align}
\label{eq:alpha}
\xi^i_t(o) &= \frac{\lambda^i_t}{\sqrt{t}(t - o +1)^{\sigma^i_t}} + \iota^i_t, \nonumber\\
\alpha^i_t(o) &= \frac{\mu_{\alpha} \xi^i_t(o)}{1 + \mu_{\alpha} \xi^i_t(o)},
\vspace{-4mm}
\end{align}
where $\mu_{\alpha}$ refers to a hyper-parameter, $t - o + 1$ represents the staleness, $t$ represents the version of the current global model, $o$ corresponds to the version of the global model that Device $i$ received before local training, $\lambda^i_t$, $\sigma^i_t$, and $\iota^i_t$ are control parameters on Device $i$ at the $t$-th global round. These three parameters are dynamically adjusted according to Formula \ref{eq:control} to reduce the loss of the global model (see details in Appendix).
\vspace{-2mm}
\begin{align}
\label{eq:control}
\lambda^i_t &= \lambda^i_{o-1} - \eta_{\lambda^i}\nabla_{\lambda^i_{o-1}} \mathcal{F}(\boldsymbol{w}_{o}), \nonumber\\
\sigma^i_t &= \sigma^i_{o-1} - \eta_{\sigma^i}\nabla_{\sigma^i_{o-1}} \mathcal{F}(\boldsymbol{w}_{o}), \\
\iota^i_t &= \iota^i_{o-1} - \eta_{\iota^i}\nabla_{\iota^i_{o-1}} \mathcal{F}(\boldsymbol{w}_{o}), \nonumber
\end{align}
where $\eta_{\lambda^i}$, $\eta_{\sigma^i}$, and $\eta_{\iota^i}$ represent the corresponding learning rates for dynamic adjustment, $\nabla_{\lambda^i_{o-1}}\mathcal{F}(\boldsymbol{w}_{o})$, $\nabla_{\sigma^i_{o-1}}\mathcal{F}(\boldsymbol{w}_{o})$, and $\nabla_{\iota^i_{o-1}}\mathcal{F}(\boldsymbol{w}_{o})$ correspond to the respective partial derivatives of the loss function. 

\begin{figure}[t]
\vspace{-3mm}
\begin{algorithm}[H]
\caption{\TheName{} on the Server}
\label{alg:serverAggregation}
\begin{algorithmic}[1]
\REQUIRE  \quad \\
$T$: The maximum number of global rounds \\
$m'$: The number of devices to be triggered \\
$\tau$: The predefined staleness limit \\
$\mathcal{T}$: The constant time period to trigger devices \\
$\boldsymbol{w}_0$: The initial global model \\
$\lambda_{0}$, $\sigma_{0}$, $\iota_{0}$: The initial control parameters\\
$\eta_{\lambda^i}$, $\eta_{\sigma^i}$, $\eta_{\iota^i}$: The learning rates for the dynamic adjustment
\ENSURE   \quad \\
$\boldsymbol{w}_{T}$: The global model at Round $T$ \\
\WHILE{The training is not finished (in parallel)} \label{alg:threadBegin}
    \IF{Should trigger new devices}
        \STATE Trigger and broadcast the global model to $m'$ devices for parallel local training
        \STATE Sleep $\mathcal{T}$
    \ENDIF
\ENDWHILE \label{alg:threadEnd}
\FOR{$t$ in $\{1, 2, ..., T\}$}
    \STATE Receive $\boldsymbol{w}^i_o$ \label{alg:receive}
    \IF{$t - o + 1 > \tau$} \label{alg:verify}
        \STATE Discard $\boldsymbol{w}^i_o$ and continue \label{alg:discard}
    \ELSE
        \STATE Update $\lambda_t^i$, $\sigma_t^i$, $\iota_t^i$ according to Formula \ref{eq:control} \label{alg:updateControl}
        \STATE Update $\alpha^i_t$ utilizing Formula \ref{eq:alpha} \label{alg:updateAlpha}
        \STATE Update $\boldsymbol{w}_t$ exploiting Formula \ref{eq:serverAggregation} \label{alg:updateModel}
    \ENDIF
\ENDFOR
\end{algorithmic}
\end{algorithm}
\vspace{-12mm}
\end{figure}

The model aggregation algorithm of \TheName{} on the server is shown in Algorithm \ref{alg:serverAggregation}. A separated thread periodically triggers $m'$ devices when the number of devices performing training is smaller than a predefined value (Lines \ref{alg:threadBegin} - \ref{alg:threadEnd}). When the server receives $\boldsymbol{w}^i_o$ (Line \ref{alg:receive}), it verifies if the uploaded model is within the staleness bound (Line \ref{alg:verify}). If not, the server ignores the $\boldsymbol{w}^i_o$ (Line \ref{alg:discard}). Otherwise, the server updates the control parameters $\lambda_t^i$, $\sigma_t^i$, $\iota_t^i$ according to Formula \ref{eq:control} (Line \ref{alg:updateControl}) and calculates $\alpha^i_t$ based on Formula \ref{eq:alpha} (Line \ref{alg:updateAlpha}). Afterward, the server updates the global model (Line \ref{alg:updateModel}). Please see the details of the convergence analysis in Appendix.

\subsection{Adaptive Model Update on Devices}
\label{subsec:MADevice}

In this subsection, we present the local training process with an adaptive local model adjustment method on devices to address Problem \ref{eq:Problem2}. 

When Device $i$ is triggered to perform local training, it receives a global model $\boldsymbol{w}_{o}$ from the server and takes it as the initial local model $\boldsymbol{w}_{o,0}$. Within the local training process, the Stochastic Gradient Descent (SGD) approach \cite{robbins1951stochastic, zinkevich2010parallelized} is exploited to update the local model based on the local dataset $\mathcal{D}_i$ as defined in Formula \ref{eq:localSGD}. 
\vspace{-1mm}
\begin{equation}
\label{eq:localSGD}
   \boldsymbol{w}_{o,l} = \boldsymbol{w}_{o,l-1} - \eta_i \nabla \mathcal{F}_{i}(\boldsymbol{w}_{o,l-1}, \zeta_{l-1}), \zeta_{l-1} \sim \mathcal{D}_i ,
\end{equation}
where $o$ is the version of the global model, $l$ represents the number of local epochs, $\eta_i$ refers to the learning rate on Device $i$, and $\nabla \mathcal{F}_{i}(\cdot)$ corresponds to the gradient based on an unbiased sampled mini-batch $\zeta_{l-1}$ from $\mathcal{D}_i$. 

In order to reduce the gap between the local model and the global model, we propose aggregating the fresh global model with the local model during the local training process of the devices. During the local training, the global model may be intensively updated simultaneously. Thus, the model aggregation with the fresh global model can well reduce the gap between the local model and the global model. However, it is complicated to determine the time slot to send the request and the weights to aggregate the fresh global model. In this section, we first propose a Reinforcement Learning (RL) method to select a proper time slot. Then, we explain the dynamic local model aggregation method. 

\subsubsection{Intelligent Time Slot Selection}

We propose an RL-based intelligent time slot selector to choose a proper time slot to request a fresh global model from the server. In order to reduce communication overhead, we assume only one fresh global model is received during the local training. When the request is sent early, the server performs few updates and the final updated local model may still suffer from severe staleness. However, when the request is sent late, the local update cannot leverage the information from the fresh global model, corresponding to inferior accuracy. Thus, it is beneficial to choose a proper time slot to send the request.

The intelligent time slot selector is composed of a meta model on the server and a local model on each device. The meta model generates an initial time slot decision for each device, and is updated when a device performs the first local training. The local model is initialized with the initial time slot and updated within the device during the following local training to generate personalized proper time slot for the fresh global model request. We exploit a Long Short-Term Memory (LSTM)-based network with a fully connected layer for the meta model and a $\mathcal{Q}$-learning method \cite{WatkinsD1992Technical} for each local model. Both the meta model and the local model generate the probability for each time slot. We exploit the $\epsilon$-greedy strategy \cite{xia2015online} to perform the selection. 

\begin{figure}[t]
\vspace{-3mm}
\begin{algorithm}[H]
\caption{\TheName{} on Devices}
\label{alg:device_passive}
\begin{algorithmic}[1]
\REQUIRE  \quad \\
$t$: The number of the meta model update \\
$t_i$: The number of local model aggregation \\
$\mathcal{L}^i$: The maximum number of epochs on Device $i$ \\
$\boldsymbol{w}_o$: The original global model with Version $o$ \\
$\boldsymbol{w}_g$: The fresh global model with Version $g$\\
$\theta^i_{t-1}$: The parameters of the meta model \\
$\gamma^i_{t_i-1}$, $\upsilon^i_{t_i-1}$: The control parameters\\
\ENSURE \quad \\
$\boldsymbol{w}^i_{o,\mathcal{L}}$: The trained local model \\
\STATE $l^* \leftarrow$ Generate a time slot using $\theta^i_{t-1}$ or $\mathcal{H}_{t_i-1}^i$ \label{alg:select}
\STATE $\boldsymbol{w}^i_{o,0} \leftarrow \boldsymbol{w}_o$ 
\FOR{$l$ in $\{1, 2, ..., \mathcal{L}^i\}$}
    \IF{$l = l^*$}
        \STATE Send a fresh global model request to server \label{alg:send}
        \STATE Receive $\boldsymbol{w}_g$ \label{alg:wait}
    \ENDIF
    \IF{$\boldsymbol{w}_g$ is updated} \label{alg:receiveFresh}
        \STATE $\beta^i_{t_i-1} \leftarrow$ Calculation based on Formula \ref{eq:betaUpdate} \label{alg:betaUpdate}
        \STATE Update $\boldsymbol{w}_{o,l-1}$ with $\beta^i_{t_i-1}$, $\boldsymbol{w}_g$ and Formula \ref{eq:deviceAggregation} \label{alg:deviceAggregation}
        \STATE Update $\gamma^i_{t_i}$ and $\upsilon^i_{t_i}$ based on Formula \ref{eq:controlDevice} \label{alg:controlDeviceUpdate}
        \STATE $\mathcal{R} \leftarrow loss_{o,l}^{b,i} - loss_{o,l}^{a,i}$ \label{alg:rewardUpdate}
        \STATE $b_{t_i} \leftarrow (1-\rho)b_{t_i-1} + \rho \mathcal{R}$  \label{alg:btUpdate}
        \STATE Update $\theta_{t}$ or $\mathcal{H}_{t_i}^i$ with $\mathcal{R}$ \label{alg:rlupdate}
    \ENDIF
    \STATE Update $\boldsymbol{w}_{o,l}$ based on Formula \ref{eq:localSGD} \label{alg:SGDUpdate}
\ENDFOR

\end{algorithmic}
\end{algorithm}
\vspace{-11mm}
\end{figure}

\begin{table*}
  \caption{The accuracy and training time with \TheName{} and diverse baseline approaches. ``Acc'' represents the convergence accuracy of the global model. ``Time'' refers to the training time to achieve a target accuracy, i.e., 0.30 for LeNet with CIFAR-10,  0.13 for LeNet with CIFAR-100, 0.40 for CNN with CIFAR-10, 0.15 for CNN with CIFAR-100, 0.25 for ResNet with CIFAR-100, and 0.12 for ResNet with Tiny-ImageNet. ``/'' represents that the method does not achieve the target accuracy.
  }
  \vspace{-3mm}
  \tiny
  \label{tab:cmp_FedASMU_1}
  \centering
  \begin{tabular}{l|ll|ll|ll|ll|ll|ll}
    \toprule
    \multirow{3}{*}{Method} & \multicolumn{4}{c|}{LeNet}  & \multicolumn{4}{c|}{CNN}  & \multicolumn{4}{c}{ResNet} \\
    \cmidrule(r){2-13} & \multicolumn{2}{c|}{ CIFAR-10}  & \multicolumn{2}{c|}{ CIFAR-100} & \multicolumn{2}{c|}{ CIFAR-10}  & \multicolumn{2}{c|}{ CIFAR-100}  & \multicolumn{2}{c|}{ CIFAR-100}  & \multicolumn{2}{c}{Tiny-ImageNet}\\
    \cmidrule(r){2-13} & Acc  & Time & Acc  & Time & Acc & Time & Acc & Time & Acc & Time & Acc & Time\\
    \midrule
    \TheName{}  &  \textbf{0.486}    & \textbf{8800}   &  \textbf{0.182}    & \textbf{20737} &  \textbf{0.603}    & \textbf{10109}   &  \textbf{0.277}    & \textbf{30569}    &  \textbf{0.358}   &  \textbf{16027}    &  \textbf{0.171}   &  \textbf{22415}       \\
    FedAvg      & 0.431   & 125514     & 0.168  & 95306           & 0.551  & 117794    & 0.243   & 73145      & 0.299  & 109680      & 0.146   & 155023     \\
    FedProx     & 0.363  & 126958    & 0.172  & 93430           & 0.371  & /        & 0.243   & 73145      & 0.302  & 109680      & 0.148   & 151935        \\
    MOON        & 0.302  & 437531    & 0.172  & 93430           & 0.47   & 100302   & 0.212   & 252703     & 0.302  & 106021      & 0.149   & 139444    \\
    FedDyn      & 0.279  & /         & 0.147  & 70260           & 0.507  & 43974    & 0.193   & 52874      & 0.328  & 73711       & 0.142   & 103661          \\
    FedAsync    & 0.478  & 36565      & 0.158  & 102113          & 0.491  & 24931    & 0.23    & 37160      & 0.315  & 21107       & 0.143   & 31288          \\
    PORT        & 0.305  & 366182    & 0.104  & /               & 0.385  & /        & 0.145   & /          & 0.314  & 35712       & 0.134   & 78155     \\
    ASO-Fed     & 0.408  & 83712     & 0.153  & 110942          & 0.482  & 92246    & 0.208   & 103090     & 0.276  & 198797      & 0.122   & 359899         \\
    FedBuff     & 0.365  & 9829      & 0.174  & 25791           & 0.364  & /        & 0.201   & 65736      & 0.315  & 27672       & 0.148   & 43523      \\
    FedSA       & 0.306  & 21077     & 0.0835 & /               & 0.508  & 20415    & 0.189   & 94169      & 0.195  & /           & 0.116  & /           \\
    \bottomrule
  \end{tabular}
  \vspace{-6mm}
\end{table*}

Within the local training process, we define the reward as the difference between the loss value before model aggregation and that after aggregation. For instance, before aggregating the fresh global model with the request sent after $l^*$ local epochs, the loss value of $\mathcal{F}_{i}(\boldsymbol{w}_{o,l^*}, \zeta_{l^*})$ is $loss_{o,l^*}^{b,i}$ and that after aggregation is $loss_{o,l^*}^{a,i}$. Then, the reward is $\mathcal{R} = loss_{o,l^*}^{b,i} - loss_{o,l^*}^{a,i}$. Inspired by \cite{Zoph2017Neural}, we update the LSTM model with Formula \ref{eq:rlupdate} once an initial aggregation is performed.
\vspace{-4mm}
\begin{equation}\label{eq:rlupdate}
    \theta_t = \theta_{t - 1} + \eta_{RL} \sum_{l = 1}^{ \mathcal{L}} \nabla_{\theta_{t - 1}} \log P(\mathcal{s}_l|\mathcal{s}_{(l-1):1}; \theta_{t - 1})(\mathcal{R} - b_t),
\end{equation}
where $\theta_t$ represents the parameters in the meta model after the $t$-th meta model update, $\eta_{RL}$ refers to the learning rate for the training process of RL, $\mathcal{L}$ is the maximum number of local epochs, $\mathcal{s}_l$ corresponds to the decision of sending the request (1) or not (0) after the $l$-th local epoch, and $b_t$ is a base value to reduce the bias of the model. The model is pre-trained with some historical data and dynamically updated during the training process of \TheName{} on each device. The $\mathcal{Q}$-learning method manages a mapping $\mathcal{H}^i$ between the decision and the reward on Device $i$, which is updated with a weighted average of historical values and reward as shown in Formula \ref{eq:qupdate}, inspired by \cite{dietterich2000hierarchical}. 
\begin{align}
\vspace{-6mm}
\label{eq:qupdate}
    &\mathcal{H}^i_{t_i}(l^*_{t_i - 1}, a_{t_i - 1})  = \mathcal{H}^i_{t_i-1}(l^*_{t_i - 1}, a_{t_i - 1}) + \phi (\mathcal{R} \nonumber \\
    &+ \psi \max_a{\mathcal{H}^i_{t_i-1}(l^*_{t_1}, a) - \mathcal{H}^i_{t_i-1}(l^*_{t_i - 1}, a_{t_i - 1})}),
\vspace{-8mm}
\end{align}
where $a_{t_i - 1}$ represents the action, $l^*_{t_i - 1}$ represents the number of local epochs to send the request within the $(t_i - 1)$-th local model aggregation, $\phi$ and $\psi$ are hyper-parameters. The action is within an action space, i.e., $a_{t_i - 1} \in \{add, stay, minus\}$, with $add$ representing adding 1 epoch to $l^*_{t_i - 1}$ ($l^*_{t_i} = l^*_{t_i - 1} + 1$), $stay$ representing staying with the same epoch ($l^*_{t_i} = l^*_{t_i - 1} + 1$), and $minus$ representing removing 1 epoch from $l^*_{t_i - 1}$ ($l^*_{t_i} = l^*_{t_i - 1} - 1$).

\subsubsection{Dynamic Local Model Aggregation}

When receiving a fresh global model $\boldsymbol{w}_g$, Device $i$ performs local model aggregation with its current local model $\boldsymbol{w}_{o,l}^b$ utilizing Formula \ref{eq:deviceAggregation}.
\vspace{-1mm}
\begin{equation}
\label{eq:deviceAggregation}
\boldsymbol{w}_{o,l}^a = (1-\beta^i_{t_i-1}) \boldsymbol{w}_{o,l}^b + \beta^i_{t_i-1} \boldsymbol{w}_g,
\vspace{-1mm}
\end{equation}
where $\beta^i_{t_i - 1}$ is the weight of the fresh global model on Device $i$ at the $(t_i - 1)$-th local global model aggregation. Formula \ref{eq:deviceAggregation} differs from Formula \ref{eq:serverAggregation} as the received fresh global model corresponds to a higher global version. We exploit Formula \ref{eq:betaUpdate} to calculate $\beta^i_{t - 1}$.
\vspace{-2mm}
\begin{align}
\label{eq:betaUpdate}
\phi^i_{t_i-1}(g, o) &= \frac{\gamma^i_{t_i-1}}{\sqrt{g}}(1-\frac{\upsilon^i_{t_i-1}}{\sqrt{g - o + 1}}), \nonumber\\
\beta^i_{t_i-1}(g, o) &= \frac{\mu_{\beta} \phi^i_{t_i-1}(g, o)}{1 + \mu_{\beta} \phi^i_{t_i-1}(g, o)},
\vspace{-4mm}
\end{align}
where $\mu_{\beta}$ is a hyper-parameter, $\gamma^i_{t_i-1}$ and $\upsilon^i_{t_i-1}$ are control parameters to be dynamically adjusted based on Formula \ref{eq:controlDevice} (see details in Appendix).
\begin{align}
\label{eq:controlDevice}
\gamma^i_{t_i} &= \gamma^i_{t_i-1} - \eta_{\gamma^i}\nabla_{\gamma^i_{t_i-1}} \mathcal{F}_i(\boldsymbol{w}_{o,l}^b, \zeta_{l-1}), \nonumber\\
\upsilon^i_{t_i} &= \upsilon^i_{t_i-1} - \eta_{\upsilon^i}\nabla_{\upsilon^i_{t_i-1}} \mathcal{F}_i(\boldsymbol{w}_{o,l}^b, \zeta_{l-1}), \zeta_{l-1} \sim \mathcal{D}_i, 
\vspace{-4mm}
\end{align}
where $\eta_{\gamma^i}$ and $\eta_{\upsilon^i}$ are learning rates for $\gamma^i_{t_i}$ and $\upsilon^i_{t_i}$.

The model update algorithm of \TheName{} on devices is shown in Algorithm \ref{alg:device_passive}. First, an epoch number $l^*$ (time slot) to send a request for a fresh global model is generated based on $\theta_{t-1}$ when $t = 1$ or $\mathcal{H}_{t_i-1}^i$ when $t\neq 1$ (Line \ref{alg:select}). In the $l^*$-th local epoch, the device sends a request to the server (Line \ref{alg:send}), and it waits for the fresh global model (Line \ref{alg:wait}). After receiving the fresh global model (Line \ref{alg:receiveFresh}), we exploit Formula \ref{eq:betaUpdate} to update $\beta^i_{t_i-1}$ (Line \ref{alg:betaUpdate}), Formula \ref{eq:deviceAggregation} to update $\boldsymbol{w}_{o,l-1}$ (Line \ref{alg:deviceAggregation}), Formula \ref{eq:controlDevice} to update $\gamma^i_{t-1}$ and $\upsilon^i_{t-1}$ (Line \ref{alg:controlDeviceUpdate}), the reward values (Line \ref{alg:rewardUpdate}), $b_{t-1}$ with $\rho$ being a hyper-parameter (Line \ref{alg:btUpdate}), $\theta_{t}$ when $t = 1$ or $\mathcal{H}_{t_i-1}^i$ when $t \neq 1$ (Line \ref{alg:rlupdate}). Finally, the local model is updated (Line \ref{alg:SGDUpdate}).

\section{Experiments}
\label{sec:exp}

In this section, we present the experimental comparison of \TheName{} with 9 state-of-the-art approaches. We first present the experimentation setup. Then, we demonstrate the experimental results.

\subsection{Experimental Setup}
\label{sec:evaluation_setup}

We consider an FL environment with a server and 100 heterogeneous devices. We consider both the asynchronous baseline approaches, i.e., FedAsync \cite{xie2019asynchronous}, PORT \cite{Su2022How}, ASO-Fed \cite{chen2020asynchronous}, FedBuff \cite{Nguyen2022FedBuff}, FedSA \cite{chen2021fedsa}, and synchronous baseline approaches, i.e., FedAvg \cite{mcmahan2017communication}, FedProx \cite{Li2020FedProx}, MOON \cite{li2021model}, and FedDyn \cite{acar2021federated}. We utilize 5 public datasets, i.e., Fashion-MNIST (FMNSIT) \cite{xiao2017fashion}, CIFAR-10 and CIFAR-100 \cite{krizhevsky2009learning}, IMDb \cite{zhou2021distilled}, and Tiny-ImageNet \cite{le2015tiny}. The data on each device is non-IID based on a Dirichlet distribution \cite{li2021federated}. We leverage 6 models to deal with the data, i.e., LeNet5 (LeNet) \cite{lecun1989handwritten}, a synthetic CNN network (CNN), ResNet20 (ResNet) \cite{He2016}, AlexNet \cite{Krizhevsky2012ImageNet}, TextCNN \cite{zhou2021distilled}, and VGG-11 (VGG) \cite{simonyan2015very}. Please see details in Appendix.

\subsection{Evaluation of \TheName{}}
\label{sec:expResults}

\begin{table*}
  \caption{The accuracy and training time with \TheName{} and diverse baseline approaches. ``Acc'' is the convergence accuracy of the global model. ``Time'' refers to the training time to achieve a target accuracy, i.e., 0.40 for AlexNet with CIFAR-10,  0.12 for AlexNet with CIFAR-100, 0.45 for VGG with CIFAR-10, 0.12 for VGG with CIFAR-100, 0.85 for TextCNN, and 0.70 for LeNet. ``/'' represents that the method does not achieve the target accuracy.
  }
  \vspace{-3mm}
  \tiny
  \label{tab:cmp_FedASMU_2}
  \centering
  \begin{tabular}{l|ll|ll|ll|ll|ll|ll}
    \toprule
    \multirow{3}{*}{Method} & \multicolumn{4}{c|}{AlexNet}  & \multicolumn{4}{c|}{VGG}  & \multicolumn{2}{c|}{TextCNN} & \multicolumn{2}{c}{LeNet} \\
    \cmidrule(r){2-13} & \multicolumn{2}{c|}{ CIFAR-10}  & \multicolumn{2}{c|}{ CIFAR-100} & \multicolumn{2}{c|}{ CIFAR-10}  & \multicolumn{2}{c|}{ CIFAR-100}  & \multicolumn{2}{c|}{IMDb} & \multicolumn{2}{c}{FMNIST} \\
    \cmidrule(r){2-13} & Acc  & Time & Acc  & Time & Acc & Time & Acc & Time & Acc & Time & Acc & Time \\
    \midrule
    \TheName{}  &  \textbf{0.490}     & \textbf{12591}   &  \textbf{0.246}    & \textbf{12150} &  \textbf{0.653}    & \textbf{43093}   &  \textbf{0.264}    & \textbf{83226}    &  \textbf{0.882}   &  \textbf{3537}      &  \textbf{0.829}   &  \textbf{8250} \\
    FedAvg      & 0.432   & 157678     & 0.205  & 92558           & 0.508  & 335866    & 0.0975   & /     & 0.874  & 13960   & 0.706  & 65000        \\
    FedProx     & 0.433  & 141125    & 0.209  & 91369          & 0.505  & 331991        & 0.0929   & /      & 0.875  & 15668    & 0.708  & 65000         \\
    MOON        & 0.429  & 157678    & 0.202  & 89297           & 0.47   & 335866   & 0.0991   & /    & 0.875  & 13960   & 0.708  & 65000       \\
    FedDyn      & 0.428  & 144999         & 0.197  & 103950           & 0.549  & 190403    & 0.218   & 307955      & 0.874  & 12674   & 0.761  & 40607             \\
    FedAsync    & 0.411  & 83693      & 0.203  & 13717          & 0.637  & 45940    & 0.147    & 375236      & 0.875  & 5837        & 0.779  & 12371         \\
    PORT        & 0.365  & /    & 0.192  & 17400               & 0.552  & 75036        & 0.209   & 120533          & 0.876  & 4884     & 0.711  & 75716         \\
    ASO-Fed     & 0.446  & 55292     & 0.238  & 60864          & 0.533  & 268349   & 0.125   & 405906     & 0.811  & /       & 0.756  & 41100        \\
    FedBuff     & 0.469  & 27763      & 0.223  & 27672           & 0.62  & 109082        & 0.238   & 167053      & 0.876  & 7671     & 0.767  & 27179        \\
    FedSA       & 0.416  & 18363       & 0.176 & 15933               & 0.383  & /    & 0.0319   & /      & 0.865  & 5251        & 0.783  & 8553            \\
    \bottomrule
  \end{tabular}
  \vspace{-6mm}
\end{table*}

As shown in Tables \ref{tab:cmp_FedASMU_1} and \ref{tab:cmp_FedASMU_2}, \TheName{} consistently corresponds to the highest convergence accuracy and training speed. Compared with synchronous baseline approaches, the training speed of \TheName{} is much faster than FedAvg (58.23\% to 92.01\%), FedProx (58.23\% to 93.06\%), MOON(74.66\% to 97.98\%), and FedDyn (42.10\% to 91.31\%) because of asynchronous model update, while the convergence accuracy of \TheName{} can still outperform the baseline approaches (0.80\% to 16.65\% higher for FedAvg, 0.70\% to 23.20\% higher for FedProx, 0.70\% to 18.30\% higher for MOON, 0.80\% to 18.90\% higher for FedDyn). Compared with asynchronous baseline approaches, \TheName{} corresponds to the fastest to achieve a target accuracy (6.19\% to 84.95\% faster than FedAsync, 27.57\% to 97.59\% faster than PORT, 70.38\% to 93.75\% faster than ASO-Fed, 10.46\% to 69.64\% faster than FedBuff, and 3.54\% to 67.5\% faster than FedSA). In addition, the accuracy of \TheName{} is significantly higher (0.70\% to 11.70\% compared with FedAsync, 0.60\% to 21.80\% compared with PORT, 2.89\% to 13.90\% compared with ASO-Fed, and 0.60\% to 23.90\% compared with FedBuff). The accuracy advantage of \TheName{} is brought by the dynamic adjustment of the weights within the model aggregation process on both the server and the devices while the high training speed is because of the asynchronous mechanism and the aggregation of the local model and the fresh global model during the local training process.

We further carry out experimental evaluation with diverse bandwidth, various device heterogeneity, and bigger number of devices (see details in Appendix). When devices have limited network connection (the bandwidth becomes modest), \TheName{} corresponds to slightly higher accuracy (5.04\% to 9.34\%) and training speed (21.21\% to 62.17\%) compared with baseline approaches. The advantages of \TheName{} become less significant due to extra global model transfer. Although \TheName{} introduces more data communication while retrieving fresh global models, it can well improve the efficiency of the FL training. When the devices are heterogeneous (the diversity of the computation and communication capacity becomes severe), \TheName{} performs much better, i.e., the advantages augment 13.67\% to 20.10\% in terms of accuracy and 85.39\% to 91.93\% in terms of efficiency. When the devices significantly differ, \TheName{} can dynamically adjust the model aggregation on both the server and devices with much better performance. The performance of \TheName{} is significantly better than that of the baseline approaches with more devices (4.52\% to 15.05\% higher in terms of accuracy and 53.47\% to 91.20\% faster), which demonstrates the excellent scalability of \TheName{}.

\begin{figure}[!t]
\centering
\subfigure[LeNet]{
\includegraphics[width=0.45\linewidth]{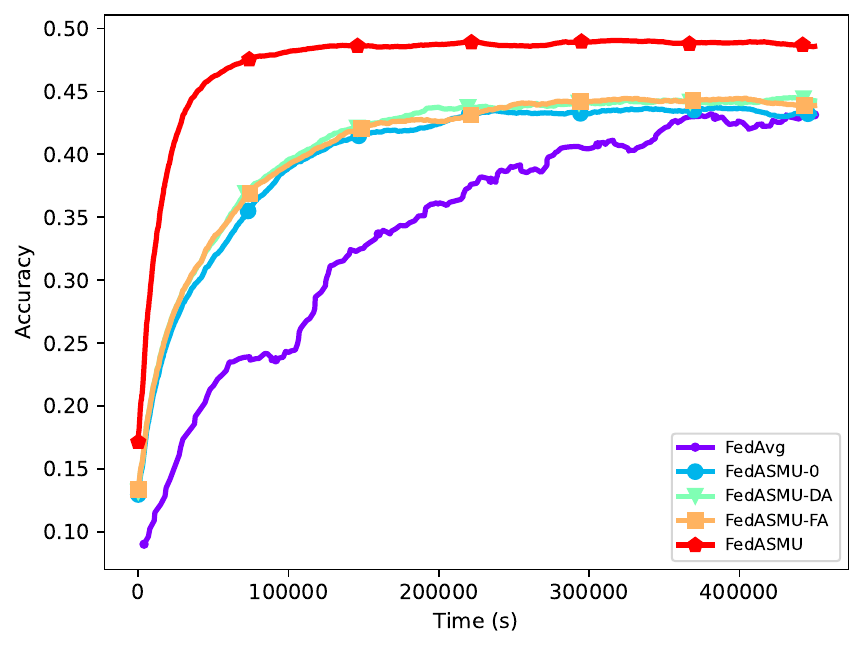}
\label{fig:abla_lenet_10}
}
\subfigure[CNN]{
\includegraphics[width=0.45\linewidth]{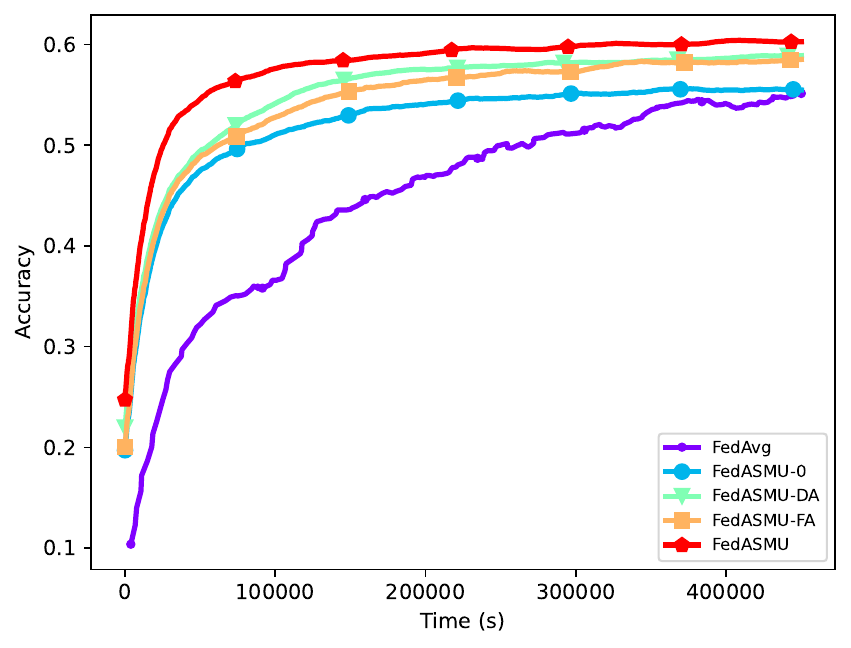}
\label{fig:abla_cnn_10}
}
\vspace{-6mm}
\caption{The accuracy and training time with \TheName{}, \TheName{}-DA, \TheName{}-FA, \TheName{}-0, and FedAvg on CIFAR-10.}
\vspace{-7mm}
\label{fig:ablation}
\end{figure}

As shown in Figure \ref{fig:ablation}, we conduct an ablation study with \TheName{}-DA, \TheName{}-FA, \TheName{}-0, and FedAvg. \TheName{}-DA represents \TheName{} without dynamic model aggregation. \TheName{}-FA refers to \TheName{} without fresh global model aggregation. \TheName{}-0 is \TheName{} without the two methods, equivalent to FedAsync with staleness bound. As the dynamic weight adjustment can improve the accuracy, \TheName{} outperforms \TheName{}-DA (1.38\% to 4.32\%) and \TheName{}-FA outperforms \TheName{}-0 (0.65\% to 3.04\%) in terms of accuracy. As the fresh global model aggregation can reduce the staleness between local models and the global model, \TheName{} corresponds to a shorter training time (44.77\% to 73.96\%) to achieve the target accuracy (0.30 for LeNet and 0.40 for CNN) and higher accuracy (1.75\% to 4.71\%) compared with \TheName{}-FA. In addition, \TheName{}-DA leads to better performance (1.04\% to 3.41\% in terms of accuracy and 15.71\% to 19.54\% faster) compared with \TheName{}-0. Both \TheName{}-DA and \TheName{}-FA outperform FedAvg in terms of accuracy (0.73\% to 3.75\%) and efficiency (72.88\% to 85.72\%). Although \TheName{}-0 corresponds to slightly higher accuracy (0.08\% to 0.34\%) compared with FedAvg, it leads to much higher efficiency (67.84\% to 82.26\% faster) because of the asynchronous mechanism. 

\section{Conclusion}
\label{sec:con}

In this paper, we propose a novel Asynchronous Stateness-aware Model Update FL framework, i.e., \TheName{}, with an asynchronous system model and two novel methods, i.e., a dynamic model aggregation method on the server and an adaptive local model adjustment method on devices. 
Extensive experimentation reveals significant advantages of \TheName{} compared with synchronous and asynchronous baseline approaches in terms of accuracy (0.60\% to 23.90\% higher) and efficiency (3.54\% to 97.98\% faster).

\bibliography{reference}

\begin{thebibliography}{73}
\providecommand{\natexlab}[1]{#1}

\bibitem[{Abad et~al.(2020)Abad, Ozfatura, Gunduz, and
  Ercetin}]{abad2020hierarchical}
Abad, M. S.~H.; Ozfatura, E.; Gunduz, D.; and Ercetin, O. 2020.
\newblock Hierarchical federated learning across heterogeneous cellular
  networks.
\newblock In \emph{IEEE Int. Conf. on Acoustics, Speech and Signal Processing
  (ICASSP)}, 8866--8870.

\bibitem[{Acar et~al.(2021)Acar, Zhao, Matas, Mattina, Whatmough, and
  Saligrama}]{acar2021federated}
Acar, D. A.~E.; Zhao, Y.; Matas, R.; Mattina, M.; Whatmough, P.; and Saligrama,
  V. 2021.
\newblock Federated Learning Based on Dynamic Regularization.
\newblock In \emph{Int. Conf. on Learning Representations ({ICLR})}, 1--36.

\bibitem[{Bard(1998)}]{Bard1998PracticalBO}
Bard, J.~F. 1998.
\newblock \emph{Practical Bilevel Optimization: Algorithms and Applications}.
\newblock Springer.

\bibitem[{Bonawitz et~al.(2019)Bonawitz, Eichner, Grieskamp, Huba, Ingerman,
  Ivanov, Kiddon, Kone{\v{c}}n{\`y}, Mazzocchi, McMahan
  et~al.}]{bonawitz2019towards}
Bonawitz, K.; Eichner, H.; Grieskamp, W.; Huba, D.; Ingerman, A.; Ivanov, V.;
  Kiddon, C.; Kone{\v{c}}n{\`y}, J.; Mazzocchi, S.; McMahan, B.; et~al. 2019.
\newblock Towards federated learning at scale: System design.
\newblock \emph{Machine Learning and Systems ({MLSys})}, 1: 374--388.

\bibitem[{Che et~al.(2023{\natexlab{a}})Che, Liu, Zhou, Ren, Zhou, Sheng, Dai,
  and Dou}]{che2023federated}
Che, T.; Liu, J.; Zhou, Y.; Ren, J.; Zhou, J.; Sheng, V.~S.; Dai, H.; and Dou,
  D. 2023{\natexlab{a}}.
\newblock Federated Learning of Large Language Models with Parameter-Efficient
  Prompt Tuning and Adaptive Optimization.
\newblock In \emph{Conf. on Empirical Methods in Natural Language Processing
  ({EMNLP})}. Singapore: Association for Computational Linguistics.

\bibitem[{Che et~al.(2022)Che, Zhang, Zhou, Zhao, Liu, Jiang, Yan, Jin, and
  Dou}]{che2022federated}
Che, T.; Zhang, Z.; Zhou, Y.; Zhao, X.; Liu, J.; Jiang, Z.; Yan, D.; Jin, R.;
  and Dou, D. 2022.
\newblock Federated Fingerprint Learning with Heterogeneous Architectures.
\newblock In \emph{IEEE Int. Conf. on Data Mining (ICDM)}, 31--40. IEEE.

\bibitem[{Che et~al.(2023{\natexlab{b}})Che, Zhou, Zhang, Lyu, Liu, Yan, Dou,
  and Huan}]{che2023fast}
Che, T.; Zhou, Y.; Zhang, Z.; Lyu, L.; Liu, J.; Yan, D.; Dou, D.; and Huan, J.
  2023{\natexlab{b}}.
\newblock Fast federated machine unlearning with nonlinear functional theory.
\newblock In \emph{Int. Conf. on Machine Learning ({ICML})}, 4241--4268. PMLR.

\bibitem[{Chen, Mao, and Ma(2021)}]{chen2021fedsa}
Chen, M.; Mao, B.; and Ma, T. 2021.
\newblock FedSA: A staleness-aware asynchronous Federated Learning algorithm
  with non-IID data.
\newblock \emph{Future Generation Computer Systems ({FGCS})}, 120: 1--12.

\bibitem[{Chen et~al.(2021)Chen, Xue, Chuai, Yang, and Liu}]{chen2021fl}
Chen, S.; Xue, D.; Chuai, G.; Yang, Q.; and Liu, Q. 2021.
\newblock FL-QSAR: a federated learning-based QSAR prototype for collaborative
  drug discovery.
\newblock \emph{Bioinformatics}, 36(22-23): 5492--5498.

\bibitem[{Chen et~al.(2020)Chen, Ning, Slawski, and
  Rangwala}]{chen2020asynchronous}
Chen, Y.; Ning, Y.; Slawski, M.; and Rangwala, H. 2020.
\newblock Asynchronous online federated learning for edge devices with non-iid
  data.
\newblock In \emph{IEEE Int. Conf. on Big Data (Big Data)}, 15--24.

\bibitem[{Dietterich(2000)}]{dietterich2000hierarchical}
Dietterich, T.~G. 2000.
\newblock Hierarchical reinforcement learning with the MAXQ value function
  decomposition.
\newblock \emph{Journal of artificial intelligence research}, 13: 227--303.

\bibitem[{Dun et~al.(2022)Dun, Garcia, Jermaine, Dimitriadis, and
  Kyrillidis}]{Dun2022EfficientAL}
Dun, C.; Garcia, M.~H.; Jermaine, C.; Dimitriadis, D.; and Kyrillidis, A. 2022.
\newblock Efficient and Light-Weight Federated Learning via Asynchronous
  Distributed Dropout.
\newblock In \emph{Int. Conf. on Artificial Intelligence and Statistics
  ({AISTATS})}.

\bibitem[{EU(2018)}]{GDPR}
EU. 2018.
\newblock {European Union's General Data Protection Regulation (GDPR)}.
\newblock \url{https://eugdpr.org/}, accessed 2018-1.

\bibitem[{He et~al.(2016)He, Zhang, Ren, and Sun}]{He2016}
He, K.; Zhang, X.; Ren, S.; and Sun, J. 2016.
\newblock Deep Residual Learning for Image Recognition.
\newblock In \emph{{IEEE} Conf. on Computer Vision and Pattern Recognition
  ({CVPR})}, 770--778. Las Vegas, USA: IEEE.

\bibitem[{Ho et~al.(2013)Ho, Cipar, Cui, Lee, Kim, Gibbons, Gibson, Ganger, and
  Xing}]{ho2013more}
Ho, Q.; Cipar, J.; Cui, H.; Lee, S.; Kim, J.~K.; Gibbons, P.~B.; Gibson, G.~A.;
  Ganger, G.; and Xing, E.~P. 2013.
\newblock More effective distributed ml via a stale synchronous parallel
  parameter server.
\newblock \emph{Advances in neural information processing systems ({NeurIPS})},
  26.

\bibitem[{Horvath et~al.(2021)Horvath, Laskaridis, Almeida, Leontiadis,
  Venieris, and Lane}]{horvath2021fjord}
Horvath, S.; Laskaridis, S.; Almeida, M.; Leontiadis, I.; Venieris, S.; and
  Lane, N. 2021.
\newblock Fjord: Fair and accurate federated learning under heterogeneous
  targets with ordered dropout.
\newblock \emph{Advances in Neural Information Processing Systems (NeurIPS)},
  34: 12876--12889.

\bibitem[{Hsu, Qi, and Brown(2019)}]{hsu2019measuring}
Hsu, T.-M.~H.; Qi, H.; and Brown, M. 2019.
\newblock Measuring the effects of non-identical data distribution for
  federated visual classification.
\newblock \emph{arXiv preprint arXiv:1909.06335}.

\bibitem[{Jia et~al.(2023)Jia, Liu, Zhou, Tian, Dong, and
  Dou}]{Jia2023Efficient}
Jia, J.; Liu, J.; Zhou, C.; Tian, H.; Dong, M.; and Dou, D. 2023.
\newblock Efficient Asynchronous Federated Learning with Sparsification and
  Quantization.
\newblock \emph{Concurrency and Computation: Practice and Experience}.
\newblock To appear.

\bibitem[{Jiang et~al.(2022)Jiang, Wang, Li, and Li}]{jiang2022pisces}
Jiang, Z.; Wang, W.; Li, B.; and Li, B. 2022.
\newblock Pisces: Efficient Federated Learning via Guided Asynchronous
  Training.
\newblock \emph{arXiv preprint arXiv:2206.09264}.

\bibitem[{Jin et~al.(2022)Jin, Ren, Zhou, Lv, Liu, and
  Dou}]{JinAccelerated2022}
Jin, J.; Ren, J.; Zhou, Y.; Lv, L.; Liu, J.; and Dou, D. 2022.
\newblock Accelerated Federated Learning with Decoupled Adaptive Optimization.
\newblock In \emph{Int. Conf. on Machine Learning ({ICML})}, volume 162,
  10298--10322.

\bibitem[{Kairouz et~al.(2021)Kairouz, McMahan, Brendan~Avent, Mehdi~Bennis,
  Bonawitz, Charles, Cormode, Cummings, D'Oliveira, Rouayheb, Evans, Gardner,
  Garrett, Gascón, Badih~Ghazi, Gruteser, Harchaoui, He, He, Huo, Hutchinson,
  Hsu, Jaggi, Javidi, Joshi, Khodak, Konečný, Korolova, Koushanfar, Koyejo,
  Lepoint, Liu, Mittal, Mohri, Nock, Özgür, Pagh, Raykova, Qi, Ramage,
  Raskar, Song, Song, Stich, Sun, Suresh, Tramèr, Vepakomma, Wang, Xiong, Xu,
  Yang, Yu, Yu, and Zhao}]{mcmahan2021advances}
Kairouz, P.; McMahan, H.~B.; Brendan~Avent, A.~B.; Mehdi~Bennis, A. N.~B.;
  Bonawitz, K.; Charles, Z.; Cormode, G.; Cummings, R.; D'Oliveira, R.~G.;
  Rouayheb, S.~E.; Evans, D.; Gardner, J.; Garrett, Z.; Gascón, A.;
  Badih~Ghazi, P. B.~G.; Gruteser, M.; Harchaoui, Z.; He, C.; He, L.; Huo, Z.;
  Hutchinson, B.; Hsu, J.; Jaggi, M.; Javidi, T.; Joshi, G.; Khodak, M.;
  Konečný, J.; Korolova, A.; Koushanfar, F.; Koyejo, S.; Lepoint, T.; Liu,
  Y.; Mittal, P.; Mohri, M.; Nock, R.; Özgür, A.; Pagh, R.; Raykova, M.; Qi,
  H.; Ramage, D.; Raskar, R.; Song, D.; Song, W.; Stich, S.~U.; Sun, Z.;
  Suresh, A.~T.; Tramèr, F.; Vepakomma, P.; Wang, J.; Xiong, L.; Xu, Z.; Yang,
  Q.; Yu, F.~X.; Yu, H.; and Zhao, S. 2021.
\newblock Advances and Open Problems in Federated Learning.
\newblock \emph{Foundations and Trends{\textregistered} in Machine Learning},
  14(1).

\bibitem[{Karimireddy et~al.(2020)Karimireddy, Kale, Mohri, Reddi, Stich, and
  Suresh}]{Karimireddy2020SCAFFOLD}
Karimireddy, S.~P.; Kale, S.; Mohri, M.; Reddi, S.; Stich, S.; and Suresh,
  A.~T. 2020.
\newblock {SCAFFOLD}: Stochastic Controlled Averaging for Federated Learning.
\newblock In \emph{Int. Conf. on Machine Learning ({ICML})}, volume 119,
  5132--5143.

\bibitem[{Khodak, Balcan, and Talwalkar(2019)}]{Khodak2019Adaptive}
Khodak, M.; Balcan, M.-F.~F.; and Talwalkar, A.~S. 2019.
\newblock Adaptive Gradient-Based Meta-Learning Methods.
\newblock In \emph{Advances in Neural Information Processing Systems
  ({NeurIPS})}, volume~32, 1--12.

\bibitem[{Koloskova, Stich, and Jaggi(2022)}]{Koloskova2022SharperCG}
Koloskova, A.; Stich, S.~U.; and Jaggi, M. 2022.
\newblock Sharper Convergence Guarantees for Asynchronous SGD for Distributed
  and Federated Learning.
\newblock \emph{ArXiv}, abs/2206.08307.

\bibitem[{Krizhevsky, Hinton et~al.(2009)}]{krizhevsky2009learning}
Krizhevsky, A.; Hinton, G.; et~al. 2009.
\newblock Learning multiple layers of features from tiny images.

\bibitem[{Krizhevsky, Sutskever, and Hinton(2012)}]{Krizhevsky2012ImageNet}
Krizhevsky, A.; Sutskever, I.; and Hinton, G.~E. 2012.
\newblock ImageNet Classification with Deep Convolutional Neural Networks.
\newblock In \emph{Annual Conf. on Neural Information Processing Systems
  ({NeurIPS})}, 1106--1114.

\bibitem[{Lai et~al.(2021)Lai, Zhu, Madhyastha, and Chowdhury}]{lai2021oort}
Lai, F.; Zhu, X.; Madhyastha, H.~V.; and Chowdhury, M. 2021.
\newblock Oort: Efficient federated learning via guided participant selection.
\newblock In \emph{{USENIX} Symposium on Operating Systems Design and
  Implementation ({OSDI})}, 19--35.

\bibitem[{Le and Yang(2015)}]{le2015tiny}
Le, Y.; and Yang, X. 2015.
\newblock Tiny imagenet visual recognition challenge.
\newblock \emph{CS 231N}, 7(7): 3.

\bibitem[{LeCun et~al.(1989)LeCun, Boser, Denker, Henderson, Howard, Hubbard,
  and Jackel}]{lecun1989handwritten}
LeCun, Y.; Boser, B.; Denker, J.; Henderson, D.; Howard, R.; Hubbard, W.; and
  Jackel, L. 1989.
\newblock Handwritten digit recognition with a back-propagation network.
\newblock In \emph{Advances in Neural Information Processing Systems
  ({NeurIPS})}, volume~2, 1--9.

\bibitem[{Li et~al.(2022)Li, Hu, Zhang, Liu, Yin, Peng, and
  Dou}]{Li2022FedHiSyn}
Li, G.; Hu, Y.; Zhang, M.; Liu, J.; Yin, Q.; Peng, Y.; and Dou, D. 2022.
\newblock FedHiSyn: A Hierarchical Synchronous Federated Learning Framework for
  Resource and Data Heterogeneity.
\newblock In \emph{Int. Conf. on Parallel Processing ({ICPP})}, 1--10.
\newblock To appear.

\bibitem[{Li, Ota, and Dong(2018)}]{li2018learning}
Li, H.; Ota, K.; and Dong, M. 2018.
\newblock Learning IoT in edge: Deep learning for the Internet of Things with
  edge computing.
\newblock \emph{IEEE network}, 32(1): 96--101.

\bibitem[{Li et~al.(2014)Li, Andersen, Park, Smola, Ahmed, Josifovski, Long,
  Shekita, and Su}]{li2014scaling}
Li, M.; Andersen, D.~G.; Park, J.~W.; Smola, A.~J.; Ahmed, A.; Josifovski, V.;
  Long, J.; Shekita, E.~J.; and Su, B.-Y. 2014.
\newblock Scaling distributed machine learning with the parameter server.
\newblock In \emph{{USENIX} Symposium on Operating Systems Design and
  Implementation ({OSDI})}, 583--598.

\bibitem[{Li et~al.(2021)Li, Diao, Chen, and He}]{li2021federated}
Li, Q.; Diao, Y.; Chen, Q.; and He, B. 2021.
\newblock Federated learning on non-iid data silos: An experimental study.
\newblock \emph{arXiv preprint arXiv:2102.02079}.

\bibitem[{Li, He, and Song(2021)}]{li2021model}
Li, Q.; He, B.; and Song, D. 2021.
\newblock Model-Contrastive Federated Learning.
\newblock In \emph{IEEE/CVF Conf. on Computer Vision and Pattern Recognition
  ({CVPR})}, 10713--10722.

\bibitem[{Li et~al.(2020)Li, Sahu, Zaheer, Sanjabi, Talwalkar, and
  Smith}]{Li2020FedProx}
Li, T.; Sahu, A.~K.; Zaheer, M.; Sanjabi, M.; Talwalkar, A.; and Smith, V.
  2020.
\newblock Federated Optimization in Heterogeneous Networks.
\newblock In \emph{Machine Learning and Systems ({MLSys})}, volume~2, 429--450.

\bibitem[{Liu et~al.(2022{\natexlab{a}})Liu, Huang, Zhou, Li, Ji, Xiong, and
  Dou}]{liu2022distributed}
Liu, J.; Huang, J.; Zhou, Y.; Li, X.; Ji, S.; Xiong, H.; and Dou, D.
  2022{\natexlab{a}}.
\newblock From distributed machine learning to federated learning: a survey.
\newblock \emph{Knowledge and Information Systems ({KAIS})}, 64(4): 885--917.

\bibitem[{Liu et~al.(2022{\natexlab{b}})Liu, Jia, Ma, Zhou, Zhou, Zhou, Dai,
  and Dou}]{liu2022multi}
Liu, J.; Jia, J.; Ma, B.; Zhou, C.; Zhou, J.; Zhou, Y.; Dai, H.; and Dou, D.
  2022{\natexlab{b}}.
\newblock Multi-Job Intelligent Scheduling With Cross-Device Federated
  Learning.
\newblock \emph{IEEE Transactions on Parallel and Distributed Systems
  ({TPDS})}, 34(2): 535--551.

\bibitem[{Liu et~al.(2023{\natexlab{a}})Liu, Wu, Yu, Ma, Feng, Zhang, Wu, Yao,
  and Dou}]{liu2021heterps}
Liu, J.; Wu, Z.; Yu, D.; Ma, Y.; Feng, D.; Zhang, M.; Wu, X.; Yao, X.; and Dou,
  D. 2023{\natexlab{a}}.
\newblock Heterps: Distributed deep learning with reinforcement learning based
  scheduling in heterogeneous environments.
\newblock \emph{Future Generation Computer Systems ({FGCS})}, 148: 106--117.

\bibitem[{Liu et~al.(2023{\natexlab{b}})Liu, Zhou, Mo, Ji, Liao, Li, Gu, and
  Dou}]{liu2023distributed}
Liu, J.; Zhou, X.; Mo, L.; Ji, S.; Liao, Y.; Li, Z.; Gu, Q.; and Dou, D.
  2023{\natexlab{b}}.
\newblock Distributed and deep vertical federated learning with big data.
\newblock \emph{Concurrency and Computation: Practice and Experience}, e7697.

\bibitem[{Liu et~al.(2021)Liu, Ho, Wang, Gao, Jin, and
  Zhang}]{liu2021federated}
Liu, M.; Ho, S.; Wang, M.; Gao, L.; Jin, Y.; and Zhang, H. 2021.
\newblock Federated learning meets natural language processing: A survey.
\newblock \emph{arXiv preprint arXiv:2107.12603}.

\bibitem[{Liu et~al.(2020)Liu, Huang, Luo, Huang, Liu, Chen, Feng, Chen, Yu,
  and Yang}]{liu2020fedvision}
Liu, Y.; Huang, A.; Luo, Y.; Huang, H.; Liu, Y.; Chen, Y.; Feng, L.; Chen, T.;
  Yu, H.; and Yang, Q. 2020.
\newblock Fedvision: An online visual object detection platform powered by
  federated learning.
\newblock In \emph{AAAI Conf. on Artificial Intelligence (AAAI)}, 13172--13179.

\bibitem[{Luo et~al.(2021)Luo, Chen, Hu, Zhang, Liang, and Feng}]{Luo2021No}
Luo, M.; Chen, F.; Hu, D.; Zhang, Y.; Liang, J.; and Feng, J. 2021.
\newblock No Fear of Heterogeneity: Classifier Calibration for Federated
  Learning with Non-IID Data.
\newblock In \emph{Advances in Neural Information Processing Systems
  ({NeurIPS})}, volume~34, 5972--5984.

\bibitem[{McMahan et~al.(2017)McMahan, Moore, Ramage, Hampson, and
  y~Arcas}]{mcmahan2017communication}
McMahan, B.; Moore, E.; Ramage, D.; Hampson, S.; and y~Arcas, B.~A. 2017.
\newblock Communication-efficient learning of deep networks from decentralized
  data.
\newblock In \emph{Artificial Intelligence and Statistics ({AISTATS})},
  1273--1282.

\bibitem[{Mitra et~al.(2021)Mitra, Jaafar, Pappas, and
  Hassani}]{Mitra2021FedLin}
Mitra, A.; Jaafar, R.; Pappas, G.~J.; and Hassani, H. 2021.
\newblock Linear Convergence in Federated Learning: Tackling Client
  Heterogeneity and Sparse Gradients.
\newblock In \emph{Advances in Neural Information Processing Systems
  ({NeurIPS})}, volume~34, 14606--14619.

\bibitem[{Nguyen et~al.(2022{\natexlab{a}})Nguyen, Pham, Pathirana, Ding,
  Seneviratne, Lin, Dobre, and Hwang}]{nguyen2022federated}
Nguyen, D.~C.; Pham, Q.-V.; Pathirana, P.~N.; Ding, M.; Seneviratne, A.; Lin,
  Z.; Dobre, O.; and Hwang, W.-J. 2022{\natexlab{a}}.
\newblock Federated learning for smart healthcare: A survey.
\newblock \emph{ACM Computing Surveys (CSUR)}, 55(3): 1--37.

\bibitem[{Nguyen et~al.(2022{\natexlab{b}})Nguyen, Malik, Zhan, Yousefpour,
  Rabbat, Malek, and Huba}]{Nguyen2022FedBuff}
Nguyen, J.; Malik, K.; Zhan, H.; Yousefpour, A.; Rabbat, M.; Malek, M.; and
  Huba, D. 2022{\natexlab{b}}.
\newblock Federated Learning with Buffered Asynchronous Aggregation.
\newblock In \emph{Int. Conf. on Artificial Intelligence and Statistics
  (AISTATS)}, volume 151, 3581--3607.

\bibitem[{Nishio and Yonetani(2019)}]{nishio2019client}
Nishio, T.; and Yonetani, R. 2019.
\newblock Client selection for federated learning with heterogeneous resources
  in mobile edge.
\newblock In \emph{IEEE Int. Conf. on communications ({ICC})}, 1--7.

\bibitem[{Ozkara et~al.(2021)Ozkara, Singh, Data, and
  Diggavi}]{ozkara2021quped}
Ozkara, K.; Singh, N.; Data, D.; and Diggavi, S. 2021.
\newblock QuPeD: Quantized Personalization via Distillation with Applications
  to Federated Learning.
\newblock In \emph{Advances in Neural Information Processing Systems
  ({NeurIPS})}, volume~34, 3622--3634.

\bibitem[{Park et~al.(2021)Park, Han, Choi, and Moon}]{park2021sageflow}
Park, J.; Han, D.-J.; Choi, M.; and Moon, J. 2021.
\newblock Sageflow: Robust federated learning against both stragglers and
  adversaries.
\newblock In \emph{Advances in Neural Information Processing Systems
  ({NeurIPS})}, volume~34, 840--851.

\bibitem[{Reddi et~al.(2021)Reddi, Charles, Zaheer, Garrett, Rush,
  Kone{\v{c}}n{\'y}, Kumar, and McMahan}]{reddi2021adaptive}
Reddi, S.~J.; Charles, Z.; Zaheer, M.; Garrett, Z.; Rush, K.;
  Kone{\v{c}}n{\'y}, J.; Kumar, S.; and McMahan, H.~B. 2021.
\newblock Adaptive Federated Optimization.
\newblock In \emph{Int. Conf. on Learning Representations ({ICLR})}, 1--38.

\bibitem[{Robbins and Monro(1951)}]{robbins1951stochastic}
Robbins, H.; and Monro, S. 1951.
\newblock A stochastic approximation method.
\newblock \emph{The annals of mathematical statistics}, 400--407.

\bibitem[{Sattler, M{\"u}ller, and Samek(2020)}]{sattler2020clustered}
Sattler, F.; M{\"u}ller, K.-R.; and Samek, W. 2020.
\newblock Clustered federated learning: Model-agnostic distributed multitask
  optimization under privacy constraints.
\newblock \emph{IEEE Transactions on Neural Networks and Learning Systems
  ({TNNLS})}, 32(8): 3710--3722.

\bibitem[{Shi, Zhou, and Niu(2020)}]{shi2020device}
Shi, W.; Zhou, S.; and Niu, Z. 2020.
\newblock Device scheduling with fast convergence for wireless federated
  learning.
\newblock In \emph{IEEE Int. Conf. on Communications (ICC)}, 1--6.

\bibitem[{Shi et~al.(2020)Shi, Zhou, Niu, Jiang, and Geng}]{shi2020joint}
Shi, W.; Zhou, S.; Niu, Z.; Jiang, M.; and Geng, L. 2020.
\newblock Joint device scheduling and resource allocation for latency
  constrained wireless federated learning.
\newblock \emph{IEEE Transactions on Wireless Communications}, 20(1): 453--467.

\bibitem[{Simonyan and Zisserman(2015)}]{simonyan2015very}
Simonyan, K.; and Zisserman, A. 2015.
\newblock Very Deep Convolutional Networks for Large-Scale Image Recognition.
\newblock In \emph{Int. Conf. on Learning Representations ({ICLR})}.

\bibitem[{Smith et~al.(2017)Smith, Chiang, Sanjabi, and
  Talwalkar}]{Smith2017Federated}
Smith, V.; Chiang, C.-K.; Sanjabi, M.; and Talwalkar, A.~S. 2017.
\newblock Federated Multi-Task Learning.
\newblock In \emph{Advances in Neural Information Processing Systems
  ({NeurIPS})}, volume~30, 1--11.

\bibitem[{Su and Li(2022)}]{Su2022How}
Su, N.; and Li, B. 2022.
\newblock How Asynchronous can Federated Learning Be?
\newblock In \emph{IEEE/ACM Int. Symposium on Quality of Service ({IWQoS})},
  1--11.

\bibitem[{Sun et~al.(2021)Sun, Huo, YANG, and Bai}]{Sun2021PartialFed}
Sun, B.; Huo, H.; YANG, Y.; and Bai, B. 2021.
\newblock PartialFed: Cross-Domain Personalized Federated Learning via Partial
  Initialization.
\newblock In \emph{Advances in Neural Information Processing Systems
  ({NeurIPS})}, volume~34, 23309--23320.

\bibitem[{Wang et~al.(2020)Wang, Liu, Liang, Joshi, and
  Poor}]{Wang2020Tackling}
Wang, J.; Liu, Q.; Liang, H.; Joshi, G.; and Poor, H.~V. 2020.
\newblock Tackling the Objective Inconsistency Problem in Heterogeneous
  Federated Optimization.
\newblock In \emph{Advances in Neural Information Processing Systems
  ({NeurIPS})}, volume~33, 7611--7623.

\bibitem[{Wang, Zhang, and Wang(2021)}]{Wang2021AsynchronousFL}
Wang, Z.; Zhang, Z.; and Wang, J. 2021.
\newblock Asynchronous Federated Learning over Wireless Communication Networks.
\newblock \emph{IEEE Int. Conf. on Communications ({ICC})}, 1--7.

\bibitem[{Watkins and Dayan(1992)}]{WatkinsD1992Technical}
Watkins, C. J. C.~H.; and Dayan, P. 1992.
\newblock Technical Note Q-Learning.
\newblock \emph{Machine Learning}, 8: 279--292.

\bibitem[{Wu et~al.(2020)Wu, He, Lin, Mao, Maple, and Jarvis}]{wu2020safa}
Wu, W.; He, L.; Lin, W.; Mao, R.; Maple, C.; and Jarvis, S. 2020.
\newblock {SAFA}: A semi-asynchronous protocol for fast federated learning with
  low overhead.
\newblock \emph{IEEE Transactions on Computers}, 70(5): 655--668.

\bibitem[{Xia and Zhao(2015)}]{xia2015online}
Xia, Z.; and Zhao, D. 2015.
\newblock Online reinforcement learning by bayesian inference.
\newblock In \emph{Int. Joint Conf. on Neural Networks (IJCNN)}, 1--6.

\bibitem[{Xiao, Rasul, and Vollgraf(2017)}]{xiao2017fashion}
Xiao, H.; Rasul, K.; and Vollgraf, R. 2017.
\newblock Fashion-mnist: a novel image dataset for benchmarking machine
  learning algorithms.
\newblock \emph{arXiv preprint arXiv:1708.07747}.

\bibitem[{Xie, Koyejo, and Gupta(2019)}]{xie2019asynchronous}
Xie, C.; Koyejo, S.; and Gupta, I. 2019.
\newblock Asynchronous federated optimization.
\newblock \emph{arXiv preprint arXiv:1903.03934}.

\bibitem[{Xu et~al.(2021)Xu, Qu, Xiang, and Gao}]{xu2021asynchronous}
Xu, C.; Qu, Y.; Xiang, Y.; and Gao, L. 2021.
\newblock Asynchronous federated learning on heterogeneous devices: A survey.
\newblock \emph{arXiv preprint arXiv:2109.04269}.

\bibitem[{Yang et~al.(2021)Yang, Wang, Xu, Chen, Bian, Liu, and
  Liu}]{yang2021characterizing}
Yang, C.; Wang, Q.; Xu, M.; Chen, Z.; Bian, K.; Liu, Y.; and Liu, X. 2021.
\newblock Characterizing impacts of heterogeneity in federated learning upon
  large-scale smartphone data.
\newblock In \emph{Web Conf. ({WWW})}, 935--946.

\bibitem[{Zhang et~al.(2022)Zhang, Liu, Jia, Zhou, and Dai}]{Zhang2022FedDUAP}
Zhang, H.; Liu, J.; Jia, J.; Zhou, Y.; and Dai, H. 2022.
\newblock {FedDUAP}: Federated Learning with Dynamic Update and Adaptive
  Pruning Using Shared Data on the Server.
\newblock In \emph{Int. Joint Conf. on Artificial Intelligence ({IJCAI})},
  2776--2782.

\bibitem[{Zhou et~al.(2022)Zhou, Liu, Jia, Zhou, Zhou, Dai, and
  Dou}]{zhou2022efficient}
Zhou, C.; Liu, J.; Jia, J.; Zhou, J.; Zhou, Y.; Dai, H.; and Dou, D. 2022.
\newblock Efficient device scheduling with multi-job federated learning.
\newblock In \emph{AAAI Conf. on Artificial Intelligence ({AAAI})}, 9971--9979.

\bibitem[{Zhou et~al.(2021{\natexlab{a}})Zhou, Tian, Zhang, Zhang, Dong, and
  Jia}]{zhou2021tea}
Zhou, C.; Tian, H.; Zhang, H.; Zhang, J.; Dong, M.; and Jia, J.
  2021{\natexlab{a}}.
\newblock TEA-fed: time-efficient asynchronous federated learning for edge
  computing.
\newblock In \emph{ACM Int. Conf. on Computing Frontiers}, 30--37.

\bibitem[{Zhou et~al.(2021{\natexlab{b}})Zhou, Pu, Ma, Li, and
  Wu}]{zhou2021distilled}
Zhou, Y.; Pu, G.; Ma, X.; Li, X.; and Wu, D. 2021{\natexlab{b}}.
\newblock Distilled One-Shot Federated Learning.
\newblock \emph{arXiv preprint arXiv:2009.07999}, abs/2009.07999(1): 1--16.

\bibitem[{Zinkevich et~al.(2010)Zinkevich, Weimer, Smola, and
  Li}]{zinkevich2010parallelized}
Zinkevich, M.; Weimer, M.; Smola, A.~J.; and Li, L. 2010.
\newblock Parallelized Stochastic Gradient Descent.
\newblock In \emph{Advances in Neural Information Processing Systems
  ({NeurIPS})}, volume~23, 1--37.

\bibitem[{Zoph and Le(2017)}]{Zoph2017Neural}
Zoph, B.; and Le, Q.~V. 2017.
\newblock Neural Architecture Search with Reinforcement Learning.
\newblock In \emph{Int. Conf. on Learning Representations ({ICLR})}.

\end{thebibliography}

\newpage

\appendix
\section{Appendix}

\subsection{Details for Model Update}

In this section, we present the details to calculate the partial deviation for the control parameters on the server and the devices.

\subsubsection{Details on the Server}

Let us denote the local model for the $(o - 1)$-th global model aggregation by $\boldsymbol{w}_{o'}^i$. Then, we get the version of the global model after aggregating the local model $\boldsymbol{w}_{o'}^i$ as $o$.
\begin{equation*}
\begin{aligned}
\nabla_{\lambda_{o-1}^i} \mathcal{F}(\boldsymbol{w}_{o}) &= (\frac{\partial \mathcal{F}(\boldsymbol{w}_{o})}{\partial \boldsymbol{w}_{o}})^\mathrm{T} \frac{\partial \boldsymbol{w}_{o}}{\partial \lambda_{o-1}^i} \\
&\approx (\frac{\partial \mathcal{F}_i(\boldsymbol{w}_{o})}{\partial \boldsymbol{w}_{o}})^\mathrm{T} \frac{\partial \boldsymbol{w}_{o}}{\partial \lambda_{o-1}^i}\\
&= g_i^\mathrm{T}(\boldsymbol{w}_{o}) \frac{\partial \boldsymbol{w}_{o}}{\partial \lambda_{o-1}^i},  
\end{aligned}
\end{equation*}
where the $\approx$ represents the approximation of the global partial deviation of $\boldsymbol{w}_{o}$ by that on Device $i$. 
\begin{equation*}
\begin{aligned}
g_i^\mathrm{T}(\boldsymbol{w}_{o}) \approx \frac{\boldsymbol{w}_{o}^i - \boldsymbol{w}_{o}}{\eta_i \mathcal{L}},
\end{aligned}
\end{equation*}
where $\boldsymbol{w}_{o}^i$ is the updated local model, and $\boldsymbol{w}_{o}$ is the original global model to generate $\boldsymbol{w}_{o}^i$. The calculation of $g_i^\mathrm{T}(\boldsymbol{w}_{o})$ does not incur extra communication. 
\begin{equation*}
\begin{aligned}
\boldsymbol{w}_{o} &= (1-\alpha^i_{o-1})\boldsymbol{w}_{o-1}+\alpha^i_{o-1}\boldsymbol{w}_{o'}^i\\
&=\boldsymbol{w}_{o-1} + \alpha^i_{o-1}(\boldsymbol{w}_{o'}^i - \boldsymbol{w}_{o-1}),
\end{aligned}
\end{equation*}
where $\boldsymbol{w}_{o-1}^i$ and $\boldsymbol{w}_{o-1}$ are independent with $\lambda^i_{o-1}$. Thus, we have:
\begin{equation*}
\begin{aligned}
\frac{\partial \boldsymbol{w}_{o}}{\partial \lambda_{o-1}^i} &= \frac{\partial(\boldsymbol{w}_{o-1} + \alpha^i_{o-1}(\boldsymbol{w}_{o'}^i - \boldsymbol{w}_{o-1}))}{\partial \lambda_{o-1}^i}\\
&= (\boldsymbol{w}_{o'}^i - \boldsymbol{w}_{o-1})\frac{\partial \alpha^i_{o-1}}{\partial \lambda_{o-1}^i}.
\end{aligned}
\end{equation*}
After elaborating $\alpha^i_{o-1}$, we have:
\begin{equation*}
\begin{aligned}
\frac{\partial \alpha^i_{o-1}}{\partial \lambda_{o-1}^i} &= \frac{\partial(1 - (1 + \mu_{\alpha} \xi^i_{o-1})^{(-1)})}{\partial \lambda_{o-1}^i}\\
&= \frac{\mu_{\alpha}}{(1+\mu_{\alpha} \xi^i_{o-1})^{2}}\frac{\partial \xi^i_{o-1}}{\partial \lambda_{o-1}^i}\\
&= \frac{\mu_{\alpha}}{(1+\mu_{\alpha} \xi^i_{o-1})^{2}}\frac{1}{\sqrt{o-1}(o - o')^{\sigma^i_{o-1}}},
\end{aligned}
\end{equation*}
where $\xi^i_{o-1}$ represents $\xi^i_{o-1}(o')$ with $o'$ representing the version of the original global model to generate updated local model $\boldsymbol{w}_{o'}^i$ at the $(o-1)$-th global round. Finally, we can calculate the partial deviation of the loss function in terms of $\lambda_{o-1}^i$:
\begin{equation*}
\begin{aligned}
\nabla_{\lambda^i_{o-1}} \mathcal{F}(\boldsymbol{w}_{o}) \approx \frac{\mu_{\alpha}(\boldsymbol{w}_{o}^i - \boldsymbol{w}_{o})(\boldsymbol{w}_{o'}^i - \boldsymbol{w}_{o-1})}{\eta_i \mathcal{L} \sqrt{o-1}(1+\mu_{\alpha} \xi^i_{o-1})^{2}(o - o')^{\sigma^i_{o-1}}}.
\end{aligned}
\end{equation*}
Similarly, we can get the partial deviation of the loss function in terms of $\sigma_{o-1}^i$ and $\iota_{o-1}^i$:
\begin{equation*}
\begin{aligned}
\nabla_{\sigma^i_{o-1}} \mathcal{F}(\boldsymbol{w}_{o}) &\approx \frac{\mu_{\alpha}\ln(\sigma^i_{o-1})(\boldsymbol{w}_{o} - \boldsymbol{w}_{o}^i)(\boldsymbol{w}_{o'}^i - \boldsymbol{w}_{o-1})}{\eta_i \mathcal{L} \sqrt{o-1}(1+\mu_{\alpha} \xi^i_{o-1})^{2}(o - o')^{\sigma^i_{o-1}}},\\
\nabla_{\iota^i_{o-1}} \mathcal{F}(\boldsymbol{w}_{o}) &\approx \frac{\mu_{\alpha}(\boldsymbol{w}_{o}^i - \boldsymbol{w}_{o})(\boldsymbol{w}_{o'}^i - \boldsymbol{w}_{o-1})}{\eta_i \mathcal{L}(1+\mu_{\alpha} \xi^i_{o-1})^{2}}.
\end{aligned}
\end{equation*}

\subsubsection{Details on the Devices}

\begin{equation*}
\begin{aligned}
\nabla_{\gamma^i_{t_i-1}} \mathcal{F}_i(\boldsymbol{w}_{o,l}^b, \zeta_{l-1}) &= (\frac{\partial \mathcal{F}_i(\boldsymbol{w}_{o,l}^b, \zeta_{l-1})}{\partial \boldsymbol{w}_{o,l}^b})^\mathrm{T} \frac{\partial \boldsymbol{w}_{o,l}^b}{\partial \gamma_{t_i-1}^i} \\
&= g_{o,l}^\mathrm{T}(\boldsymbol{w}_{o,l}^b) \frac{\partial \boldsymbol{w}_{o,l}^b}{\partial \gamma_{t_i-1}^i},  \\
\end{aligned}
\end{equation*}
where $g_{o,l}^\mathrm{T}(\boldsymbol{w}_{o,l}^b)$ is the local gradient on Device $i$ with $\boldsymbol{w}_{o,l}^b$ and $\zeta_{l-1}$. 
\begin{equation*}
\begin{aligned}
\boldsymbol{w}_{o,l}^b &= (1-\beta^i_{t_i-1}) \boldsymbol{w}_{o,l}^a + \beta^i_{t_i-1} \boldsymbol{w}_g \\
&= \boldsymbol{w}_{o,l}^a + \beta^i_{t_i-1}(\boldsymbol{w}_g - \boldsymbol{w}_{o,l}^a),
\end{aligned}
\end{equation*}
where $\boldsymbol{w}_g$ and $\boldsymbol{w}_{o,l}^a$ are independent with $\gamma_{t_i-1}^i$. Then, we have:
\begin{equation*}
\begin{aligned}
\frac{\partial \boldsymbol{w}_{o,l}^b}{\partial \gamma_{t_i-1}^i} &= \frac{\partial(\boldsymbol{w}_{o,l}^a + \beta^i_{t_i-1}(\boldsymbol{w}_g - \boldsymbol{w}_{o,l}^a))}{\partial \gamma_{t_i-1}^i} \\
&= (\boldsymbol{w}_g - \boldsymbol{w}_{o,l}^a)\frac{\partial \beta^i_{t_i-1}}{\partial \gamma_{t_i-1}^i}.
\end{aligned}
\end{equation*}
After elaborating $\beta^i_{t_i-1}$, we have:
\begin{equation*}
\begin{aligned}
\frac{\partial \beta^i_{t_i-1}}{\partial \gamma_{t_i-1}^i} &= \frac{\partial(1-(1 + \mu_{\beta} \phi^i_{t_i-1})^{-1})}{\partial \gamma_{t_i-1}^i} \\
&= \frac{\mu_{\beta}}{(1 + \mu_{\beta} \phi^i_{t_i-1})^{2}}\frac{\partial \phi^i_{t_i-1}}{\partial \gamma_{t_i-1}^i} \\
&= \frac{\mu_{\beta}}{\sqrt{g}(1 + \mu_{\beta} \phi^i_{t_i-1})^{2}}(1-\frac{\upsilon^i_{t_i-1}}{\sqrt{g - o + 1}}),
\end{aligned}
\end{equation*}
where $\phi^i_{t_i-1}$ represents $\phi^i_{t_i-1}(g,o)$. Finally, we can calculate the partial deviation of the loss function in terms of $\gamma_{t_i-1}^i$:
\begin{equation*}
\begin{aligned}
&\nabla_{\gamma^i_{t_i-1}} \mathcal{F}_i(\boldsymbol{w}_{o,l}^b, \zeta_{l-1}) \\
= ~&(\boldsymbol{w}_g - \boldsymbol{w}_{o,l}^a)\frac{\mu_{\beta}g_{o,l}^\mathrm{T}(\boldsymbol{w}_{o,l}^b)}{\sqrt{g}(1 + \mu_{\beta} \phi^i_{t_i-1})^{2}}(1-\frac{\upsilon^i_{t_i-1}}{\sqrt{g - o + 1}}).
\end{aligned}
\end{equation*}
Similarly, we can get the partial deviation of the loss function in terms of $\upsilon^i_{t_i-1}$:
\begin{equation*}
\begin{aligned}
\nabla_{\upsilon^i_{t_i-1}} \mathcal{F}_i(\boldsymbol{w}_{o,l}^b, \zeta_{l-1}) &= \frac{\mu_{\beta}\gamma^i_{t_i-1}g_{o,l}^\mathrm{T}(\boldsymbol{w}_{o,l}^b)(\boldsymbol{w}_{o,l}^a - \boldsymbol{w}_g)}{\sqrt{g}\sqrt{g - o + 1}(1 + \mu_{\beta} \phi^i_{t_i-1})^{2}}.
\end{aligned}
\end{equation*}

\subsection{Convergence Analysis}

In this section, we present the assumptions, the convergence guarantees of \TheName{}, and the proof.
\begin{assumption} 
\label{ass:smooth}
($L$-smoothness) The loss function $\mathcal{F}_i$ is differentiable and $L$-smooth for each device $i \in \mathcal{M}$ and $\forall x, y$, 
$
\mathcal{F}_i(y) - \mathcal{F}_i(x) \leq \ip{\nabla \mathcal{F}_i(x)}{y-x} + \frac{L}{2} \|y-x\|^2
$
with $L > 0$.
\end{assumption}
\begin{assumption} 
\label{ass:convex}
($\mu$-strongly convex)
The loss function $\mathcal{F}_i$ is $\mu$-strongly convex for each device $i \in \mathcal{M}$: $ \langle \nabla \mathcal{F}_i(x) - \nabla \mathcal{F}_i(y), x - y \rangle  \geq  \mu \parallel x - y \parallel ^ 2$ with $\mu > 0$.
\end{assumption}
\begin{assumption} 
\label{ass:sampling}
(Unbiased sampling)
The local sampling is unbiased and the local gradients are unbiased stochastic gradients $\mathbb{E}_{\zeta_l \sim \mathcal{D}_i} [\nabla \mathcal{F}_i(\boldsymbol{w}_{o,l}; \zeta_l)] = \nabla \mathcal{F}_i(\boldsymbol{w}_{o,l})$.
\end{assumption}
\begin{assumption} 
\label{ass:gradient}
(Bounded local gradient)
The stochastic gradients are bounded on each device $i \in \mathcal{M}$: $\mathbb{E}_{\zeta_l \sim \mathcal{D}_i} \parallel \nabla \mathcal{F}_i(\boldsymbol{w}_{o,l}; \zeta_l) \parallel^2 \le{\mathcal{G}^2}$.
\end{assumption}
\begin{assumption} 
\label{ass:variance}
(Bounded local variance)
The variance of local stochastic gradients are bounded on each device $i \in \mathcal{M}$ is bounded: $\mathbb{E}_{\zeta_l \sim \mathcal{D}_i} \parallel \nabla \mathcal{F}_i(\boldsymbol{w}_{o,l}; \zeta_l) - \mathcal{F}(\boldsymbol{w}_{o,l}) \parallel^2 \le{\mathcal{V}^2}$.
\end{assumption}
\begin{theorem}
Let Assumptions \ref{ass:smooth} - \ref{ass:variance} hold, after $T$ global updates, \TheName{} converges to a critical point:
\begin{align*}
&\min_{t=0}^{T} \E\parallel \nabla \mathcal{F}(\boldsymbol{w}_{o,l}) \parallel^2 \\
\leq~& \frac{2 \E\left[ \mathcal{F}(\boldsymbol{w}_{0}) - \mathcal{F}(\boldsymbol{w}_{T}) \right]}{\alpha_{min} \mathcal{L}_{min}^3} + \OM\left( \frac{L \mathcal{G}^2 \mathcal{L}_{max}}{\mathcal{L}_{min}^3} \right) \\
&\quad + \OM\left( \frac{\mathcal{L}^i \mathcal{V}^2}{ \mathcal{L}_{min}^7}\right) + \OM\left( \frac{\tau \mathcal{G}^2 \mathcal{L}_{max}}{ \mathcal{L}_{min}^7} \right) \\
&\quad + \OM\left( \frac{ \mathcal{G}^2 \mathcal{L}_{max} }{\mathcal{L}_{min}^3} \right) + \OM\left( \frac{ L \mathcal{G}^2 \mathcal{L}_{max} }{\mathcal{L}_{min}^3} \right)\\
&  + \OM\left(\frac{ L \tau^2 \mathcal{G}^2 \mathcal{L}_{max}^2}{\mathcal{L}_{min}^3} \right) + \OM\left( \frac{ L \tau^2 \mathcal{G}^2 \mathcal{L}_{max}^2 }{\mathcal{L}_{min}^3} \right),
\end{align*}
where $\alpha_{min} \leq \alpha_t^i$, $\mathcal{L}_{min} \leq \mathcal{L}_t \leq \mathcal{L}_{max}$, $\eta_i = \frac{1}{\sqrt{T}}$, $\forall i \in \mathcal{M}$, and $T = \mathcal{L}_{min}^6$.
\end{theorem}

\begin{proof}
First, we denote the optimal model by $\boldsymbol{w}^*$, the new fresh global gradient is not received at the $l$-th local epoch, and we have the following inequality with the vanilla SGD in devices: 
\begin{align*}
&~\E\left[ \mathcal{F}(\boldsymbol{w}_{o,l}) - \mathcal{F}(\boldsymbol{w}^*) \right] \\
= &~\E_{\zeta_{l-1} \sim \mathcal{D}_i}\left[ \mathcal{F}(\boldsymbol{w}_{o,l-1} - \eta_i \nabla \mathcal{F}_{i}(\boldsymbol{w}_{o,l-1}, \zeta_{l-1})) - \mathcal{F}(\boldsymbol{w}^*) \right]\\
\leq &~\mathcal{F}(\boldsymbol{w}_{o,l-1}) - F(\boldsymbol{w}^*) \\
&\quad - \eta_i \E_{\zeta_{l-1} \sim \mathcal{D}_i}\left[  \ip{\nabla \mathcal{F}(\boldsymbol{w}_{o,l-1})}{\nabla \mathcal{F}_{i}(\boldsymbol{w}_{o,l-1}, \zeta_{l-1})} \right] \\
&\quad + \frac{L \eta_i^2}{2} \E_{\zeta_{l-1} \sim \mathcal{D}_i}\left[  \| \nabla \mathcal{F}_{i}(\boldsymbol{w}_{o,l-1}, \zeta_{l-1}) \|^2 \right]\\
\leq &~\mathcal{F}(\boldsymbol{w}_{o,l-1}) - F(\boldsymbol{w}^*)  + \frac{L \eta_i^2 \mathcal{G}^2}{2}\\
&\quad - \eta_i \underbrace{\E_{\zeta_{l-1} \sim \mathcal{D}_i}\left[  \ip{\nabla \mathcal{F}(\boldsymbol{w}_{o,l-1})}{\nabla \mathcal{F}_{i}(\boldsymbol{w}_{o,l-1}, \zeta_{l-1})} \right]}_{A}, \tag{1}\label{eq1}
\end{align*}
where the first inequality comes from $L$-smoothness and the second one is from bounded local gradient. Then, we focus on $A$. 
\begin{align*}
&~\mathbb{E}_{\zeta_{l-1} \sim \mathcal{D}_i} \parallel \nabla \mathcal{F}_i(\boldsymbol{w}_{o,l-1}; \zeta_{l-1}) - \nabla \mathcal{F}(\boldsymbol{w}_{o,l-1}) \parallel^2 \\
= &~\E\parallel \nabla \mathcal{F}(\boldsymbol{w}_{o,l-1}) \parallel^2  - 2A\\
&\quad + ~\mathbb{E}_{\zeta_{l-1} \sim \mathcal{D}_i} \parallel \nabla \mathcal{F}_i(\boldsymbol{w}_{o,l-1}; \zeta_{l-1}) \parallel^2. 
\end{align*}
Based on the bounded local variance assumption, we have:
\begin{align*}
\mathcal{V}^2
= ~&\E\parallel \nabla \mathcal{F}(\boldsymbol{w}_{o,l-1}) \parallel^2 - 2A \\
&\quad + ~\mathbb{E}_{\zeta_{l-1} \sim \mathcal{D}_i} \parallel \nabla \mathcal{F}_i(\boldsymbol{w}_{o,l-1}; \zeta_{l-1}) \parallel^2,
\end{align*}
and we can get $A$:
\begin{align*}
A = &\frac{1}{2} (\E\parallel \nabla \mathcal{F}(\boldsymbol{w}_{o,l-1}) \parallel^2 - ~\mathcal{V}^2 \\
&\quad + ~\mathbb{E}_{\zeta_{l-1} \sim \mathcal{D}_i} \parallel \nabla \mathcal{F}_i(\boldsymbol{w}_{o,l-1}; \zeta_{l-1}) \parallel^2),
\end{align*}
Plug this into Formula \ref{eq1}, and we have:
\begin{align*}
&~\E\left[ \mathcal{F}(\boldsymbol{w}_{o,l}) - \mathcal{F}(\boldsymbol{w}^*) \right] \\
\leq &~\mathcal{F}(\boldsymbol{w}_{o,l-1}) - F(\boldsymbol{w}^*) + \frac{L \eta_i^2 \mathcal{G}^2}{2}\\
&\quad - \frac{\eta_i}{2} (\E\parallel \nabla \mathcal{F}(\boldsymbol{w}_{o,l-1}) \parallel^2 - \mathcal{V}^2\\
&\quad + ~\mathbb{E}_{\zeta_{l-1} \sim \mathcal{D}_i} \parallel \nabla \mathcal{F}_i(\boldsymbol{w}_{o,l-1}; \zeta_{l-1}) \parallel^2 ) \\
\leq &~\mathcal{F}(\boldsymbol{w}_{o,l-1}) - F(\boldsymbol{w}^*) - \frac{\eta_i}{2} \E\parallel \nabla \mathcal{F}(\boldsymbol{w}_{o,l-1}) \parallel^2 \\
&\quad + \frac{L \eta_i^2 \mathcal{G}^2 + \eta_i \mathcal{V}^2}{2},
\end{align*}
where the second inequality is because $\mathbb{E}_{\zeta_{l-1} \sim \mathcal{D}_i} \parallel \nabla \mathcal{F}_i(\boldsymbol{w}_{o,l-1}; \zeta_{l-1}) \parallel^2 \geq 0$. By rearranging the terms and telescoping, we have:
\begin{align*}
\E\parallel \nabla \mathcal{F}(\boldsymbol{w}_{o,l-1}) \parallel^2 \leq &\frac{2}{\eta_i} \E\left[ \mathcal{F}(\boldsymbol{w}_{o,l-1}) - \mathcal{F}(\boldsymbol{w}_{o,l}) \right] \\
&\quad + L \eta_i \mathcal{G}^2 + \mathcal{V}^2.
\end{align*}
However, when the fresh global model $\boldsymbol{w}_g$ is received right at the $l^*$-th local epoch, we have:
\begin{align*}
&\E\parallel \nabla \mathcal{F}(\boldsymbol{w}_{o,l^*}) \parallel^2 \\
\leq &\frac{2}{\eta_i} \E\left[ \mathcal{F}(\boldsymbol{w}_{o,l^*-1}^a) - \mathcal{F}(\boldsymbol{w}_{o,l^*}) \right] + L \eta_i \mathcal{G}^2 + \mathcal{V}^2 \\
= &\frac{2}{\eta_i} \E\left[ \mathcal{F}((1-\beta^i_{t_i-1}) \boldsymbol{w}_{o,l^*-1}^b + \beta^i_{t_i-1} \boldsymbol{w}_g) - \mathcal{F}(\boldsymbol{w}_{o,l^*}) \right] \\
&\quad + L \eta_i \mathcal{G}^2 + \mathcal{V}^2 \\
\leq &\frac{2}{\eta_i} \E\left[ (1-\beta^i_{t_i-1}) \mathcal{F}(\boldsymbol{w}_{o,l^*-1}^b) + \beta^i_{t_i-1} \mathcal{F}(\boldsymbol{w}_g) -  \mathcal{F}(\boldsymbol{w}_{o,l^*}) \right] \\
&\quad + L \eta_i \mathcal{G}^2 + \mathcal{V}^2 \\
= &\frac{2}{\eta_i} \E\left[ (1-\beta^i_{t_i-1}) \mathcal{F}(\boldsymbol{w}_{o,l^*-1}) + \beta^i_{t_i-1} \mathcal{F}(\boldsymbol{w}_g) - \mathcal{F}(\boldsymbol{w}_{o,l^*}) \right] \\
&\quad + L \eta_i \mathcal{G}^2 + \mathcal{V}^2 \\
= &\frac{2}{\eta_i} \E\left[ \mathcal{F}(\boldsymbol{w}_{o,l^*-1}) - \mathcal{F}(\boldsymbol{w}_{o,l^*}) \right]\\
&\quad + \frac{2 \beta^i_{t_i-1}}{\eta_i} \E\left[ \mathcal{F}(\boldsymbol{w}_g) -\mathcal{F}(\boldsymbol{w}_{o,l^*-1}) \right] \\
&\quad + L \eta_i \mathcal{G}^2 + \mathcal{V}^2 \tag{2}\label{eq2}
\end{align*}
where the second inequality is because of convexity of $\mathcal{F}(\cdot)$. Then, we can get:
\begin{align*}
&~\sum_{l=1}^{\mathcal{L}^i} \E\parallel \nabla \mathcal{F}(\boldsymbol{w}_{o,l}) \parallel^2 \\
\leq &~\frac{2}{\eta_i} \E\left[ \mathcal{F}(\boldsymbol{w}_{o,0}) - \mathcal{F}(\boldsymbol{w}_{o,\mathcal{L}^i}) \right] + \mathcal{L}^i (L \eta_i \mathcal{G}^2 + \mathcal{V}^2)\\
&\quad + \frac{2 \beta^i_{t_i-1}}{\eta_i} \E\left[ \mathcal{F}(\boldsymbol{w}_g) -\mathcal{F}(\boldsymbol{w}_{o,l^*-1}) \right] \\
= &~\frac{2}{\eta_i} \E\left[ \mathcal{F}(\boldsymbol{w}_o) - \mathcal{F}(\boldsymbol{w}_o^i)  \right] + \mathcal{L}^i (L \eta_i \mathcal{G}^2 + \mathcal{V}^2)\\
&\quad + \frac{2 \beta^i_{t_i-1}}{\eta_i} \E\left[ \mathcal{F}(\boldsymbol{w}_g) -\mathcal{F}(\boldsymbol{w}_{o,l^*-1}) \right] \\
= &~\frac{2}{\eta_i} \underbrace{\E\left[ \mathcal{F}(\boldsymbol{w}_o) - \mathcal{F}(\boldsymbol{w}_o^i)\right]}_{B} + \mathcal{L}^i (L \eta_i \mathcal{G}^2 + \mathcal{V}^2)\\
&\quad + \frac{2}{\eta_i} \beta^i_{t_i-1}\underbrace{\E\left[ \mathcal{F}(\boldsymbol{w}_{o,0}) - \mathcal{F}(\boldsymbol{w}_{o,l^*-1}) \right]}_{C} \\
&\quad + \frac{2}{\eta_i} \beta^i_{t_i-1}\underbrace{\E\left[ \mathcal{F}(\boldsymbol{w}_g) - \mathcal{F}(\boldsymbol{w}_o) \right]}_{D}.
\end{align*}
First, we focus on the calculation of $B$. 
\begin{align*}
&\E\left[ \mathcal{F}(\boldsymbol{w}_{t+1}) - \mathcal{F}(\boldsymbol{w}_{t}) \right] \\
= &~\E\left[ \mathcal{F}((1-\alpha^i_t) \boldsymbol{w}_{t} + \alpha^i_t \boldsymbol{w}_o^i) - \mathcal{F}(\boldsymbol{w}_{t}) \right] \\
\leq &~\E\left[ (1-\alpha^i_t) \mathcal{F}(\boldsymbol{w}_{t}) + \alpha^i_t \mathcal{F}(\boldsymbol{w}_o^i) - \mathcal{F}(\boldsymbol{w}_{t}) \right] \\
= &~\alpha^i_t \E\left[ \mathcal{F}(\boldsymbol{w}_o^i)  - \mathcal{F}(\boldsymbol{w}_{t}) \right] \\
= &~\alpha^i_t \E\left[ \mathcal{F}(\boldsymbol{w}_o^i) - \mathcal{F}(\boldsymbol{w}_o) + \mathcal{F}(\boldsymbol{w}_o)  - \mathcal{F}(\boldsymbol{w}_{t}) \right],
\end{align*}
where the inequility is because $\mathcal{F}(\cdot)$ is convex. Then, we have:
\begin{align*}
&~\E\left[ \mathcal{F}(\boldsymbol{w}_{t+1}) - \mathcal{F}(\boldsymbol{w}_{t}) \right] \\
\leq &~\alpha^i_t \E\left[ \mathcal{F}(\boldsymbol{w}_o^i) - \mathcal{F}(\boldsymbol{w}_o) + \mathcal{F}(\boldsymbol{w}_o)  - \mathcal{F}(\boldsymbol{w}_{t}) \right].
\end{align*}
And, we can get: 
\begin{align*}
&~\E\left[ \mathcal{F}(\boldsymbol{w}_o) - \mathcal{F}(\boldsymbol{w}_o^i) \right] \\
\leq &~\frac{1}{\alpha^i_t} \E\left[ \mathcal{F}(\boldsymbol{w}_{t}) - \mathcal{F}(\boldsymbol{w}_{t+1}) \right] + \E\left[ \mathcal{F}(\boldsymbol{w}_o)  - \mathcal{F}(\boldsymbol{w}_{t}) \right] .
\end{align*}
Using $L$-smoothness, we have:
\begin{align*}
&~\E\left[ \mathcal{F}(\boldsymbol{w}_o)  - \mathcal{F}(\boldsymbol{w}_t) \right] \\
\leq &~\ip{\nabla \mathcal{F}(\boldsymbol{w}_t)}{\boldsymbol{w}_o - \boldsymbol{w}_t} + \frac{L}{2}\parallel \boldsymbol{w}_o - \boldsymbol{w}_t \parallel ^2 \\
\leq &~\parallel \nabla \mathcal{F}(\boldsymbol{w}_t) \parallel \parallel \boldsymbol{w}_o - \boldsymbol{w}_t \parallel + \frac{L}{2}\parallel \boldsymbol{w}_o - \boldsymbol{w}_t \parallel ^2
\end{align*}
As the fresh global model is incurred to reduce the difference between the local model and the global model, the difference between the global models of two versions is because of the local updates. Then, we have the upper bound of local updates:
\begin{align*}
&~\parallel \boldsymbol{w}_{o,0}  - \boldsymbol{w}_{o,\mathcal{L}^i} \parallel \\
\leq &~\parallel \boldsymbol{w}_{o,0}  - \boldsymbol{w}_{o,1} \parallel + \parallel \boldsymbol{w}_{o,1}  - \boldsymbol{w}_{o,2} \parallel + \cdots \\
& \quad + \parallel \boldsymbol{w}_{o,\mathcal{L}^i - 1}  - \boldsymbol{w}_{o,\mathcal{L}^i} \parallel \\
\leq &~\eta_i \mathcal{L}^i \mathcal{G}.
\end{align*}
And, we get:
\begin{align*}
\parallel \boldsymbol{w}_{o} - \boldsymbol{w}_{o+1} \parallel &= \parallel \boldsymbol{w}_{o} - (1-\alpha_t^i) \boldsymbol{w}_{o} - \alpha_t^i \boldsymbol{w}_{o,\mathcal{L}^i} \parallel \\
&= \alpha_t^i \parallel \boldsymbol{w}_{o,0} - \boldsymbol{w}_{o,\mathcal{L}^i} \parallel \\
&\leq \eta_i \alpha_t^i \mathcal{L} \mathcal{G}.
\end{align*}
Thus, we have:
\begin{align*}
\parallel \boldsymbol{w}_o - \boldsymbol{w}_t \parallel &\leq (t - o + 1) \eta_i \alpha_t^i \mathcal{L}^i \mathcal{G},
\end{align*}
where $t - o + 1\leq \tau$ because of staleness bound. Then, we can get:
\begin{align*}
\parallel \boldsymbol{w}_o - \boldsymbol{w}_t \parallel &\leq \tau \eta_i \alpha_t^i \mathcal{L}^i \mathcal{G}.
\end{align*}
Then, we have:
\begin{align*}
&~\E\left[ \mathcal{F}(\boldsymbol{w}_o)  - \mathcal{F}(\boldsymbol{w}_t) \right] \\
\leq &~\parallel \nabla \mathcal{F}(\boldsymbol{w}_t) \parallel \parallel \boldsymbol{w}_o - \boldsymbol{w}_t \parallel + \frac{L}{2}\parallel \boldsymbol{w}_o - \boldsymbol{w}_t \parallel ^2 \\
&\leq \tau \eta_i \alpha_t^i \mathcal{L}^i \mathcal{G}^2 + \frac{L}{2} (\tau \eta_i \alpha_t^i \mathcal{L}^i \mathcal{G})^2.
\end{align*}
And, we can calculate $B$:
\begin{align*}
B \leq &\frac{1}{\alpha^i_t} \E\left[ \mathcal{F}(\boldsymbol{w}_{t}) - \mathcal{F}(\boldsymbol{w}_{t+1}) \right] + \tau \eta_i \alpha_t^i \mathcal{L} \mathcal{G}^2 \\
&\quad + \frac{L}{2} (\tau \eta_i \alpha_t^i \mathcal{L} \mathcal{G})^2.
\end{align*}
Now, we focus on the calculation of $C$. Based on the convexity of $\mathcal{F}(\cdot)$, we have:
\begin{align*}
&~\E\left[ \mathcal{F}(\boldsymbol{w}_{o,l-1}) - \mathcal{F}(\boldsymbol{w}_{o,l}) \right] \\
\leq &~\ip{\nabla \mathcal{F}(\boldsymbol{w}_{o,l})}{\boldsymbol{w}_{o,l-1} - \boldsymbol{w}_{o,l}} + \frac{L}{2}\parallel \boldsymbol{w}_{o,l-1} - \boldsymbol{w}_{o,l} \parallel ^2 \\
= &~\eta_i \ip{\nabla \mathcal{F}(\boldsymbol{w}_{o,l})}{\nabla \mathcal{F}_i(\boldsymbol{w}_{o,l-1})} + \frac{L\eta_i^2}{2}\parallel \nabla \mathcal{F}_i(\boldsymbol{w}_{o,l-1}) \parallel ^2 \\
\leq &~\frac{\eta_i}{2} (\parallel \nabla \mathcal{F}(\boldsymbol{w}_{o,l}) \parallel^2 + \parallel \nabla \mathcal{F}_i(\boldsymbol{w}_{o,l-1}) \parallel^2) + \frac{L \eta_i^2 \mathcal{G}^2}{2} \\
\leq &~ \frac{2 \eta_i  + L \eta_i^2}{2} \mathcal{G}^2,
\end{align*}
Then, we can have:
\begin{align*}
C = &~\E\left[ \mathcal{F}(\boldsymbol{w}_{o,0}) - \mathcal{F}(\boldsymbol{w}_{o,l^*-1}) \right] \\
\leq &~\frac{2 \eta_i  + L \eta_i^2}{2} (l^*-1) \mathcal{G}^2 \\
\leq &~\frac{2 \eta_i  + L \eta_i^2}{2} \mathcal{L}^i \mathcal{G}^2.
\end{align*}
Next, we focus on the calculation of $D$. 
\begin{align*}
D &\leq \ip{\nabla \mathcal{F}(\boldsymbol{w}_o)}{\boldsymbol{w}_g - \boldsymbol{w}_o} + \frac{L}{2} \|\boldsymbol{w}_g - \boldsymbol{w}_o\|^2 \\
&\leq \|\nabla \mathcal{F}(\boldsymbol{w}_o)\| \|\boldsymbol{w}_g - \boldsymbol{w}_o\| + \frac{L}{2} \|\boldsymbol{w}_g - \boldsymbol{w}_o\|^2. 
\end{align*}
As $(g - o \leq \tau)$ because of staleness bound, we have:
\begin{align*}
\parallel \boldsymbol{w}_g - \boldsymbol{w}_o \parallel \leq (g - o) \eta_i \alpha_t^i \mathcal{L} \mathcal{G} \leq \tau \eta_i \alpha_t^i \mathcal{L} \mathcal{G}.
\end{align*}
Then, we have:
\begin{align*}
D &\leq \|\nabla \mathcal{F}(\boldsymbol{w}_o)\| \|\boldsymbol{w}_g - \boldsymbol{w}_o\| + \frac{L}{2} \|\boldsymbol{w}_g - \boldsymbol{w}_o\|^2 \\
&\leq \tau \eta_i \alpha_t^i \mathcal{L} \mathcal{G}^2 + \frac{L}{2} (\tau \eta_i \alpha_t^i \mathcal{L} \mathcal{G})^2.
\end{align*}
By rearranging the terms, we have
\begin{align*}
&~\sum_{l=1}^{\mathcal{L}^i} \E\parallel \nabla \mathcal{F}(\boldsymbol{w}_{o,l}) \parallel^2 \\
\leq &~\frac{2}{\eta_i} \underbrace{\E\left[ \mathcal{F}(\boldsymbol{w}_o) - \mathcal{F}(\boldsymbol{w}_o^i)\right]}_{B} \\
&\quad + \frac{2}{\eta_i} \beta^i_{t_i-1}\underbrace{\E\left[ \mathcal{F}(\boldsymbol{w}_{o,0}) - \mathcal{F}(\boldsymbol{w}_{o,l^*-1}) \right]}_{C} \\
&\quad + \frac{2}{\eta_i} \beta^i_{t_i-1}\underbrace{\E\left[ \mathcal{F}(\boldsymbol{w}_g) - \mathcal{F}(\boldsymbol{w}_o) \right]}_{D} + \mathcal{L}^i (L \eta_i \mathcal{G}^2 + \mathcal{V}^2) \\
\leq & ~\frac{2}{\eta_i} (\frac{1}{\alpha^i_t} \E\left[ \mathcal{F}(\boldsymbol{w}_{t}) - \mathcal{F}(\boldsymbol{w}_{t+1}) \right]) \\
&\quad + \frac{2}{\eta_i} (\tau \eta_i \alpha_t^i \mathcal{L}^i \mathcal{G}^2 + \frac{L}{2} (\tau \eta_i \alpha_t^i \mathcal{L}^i \mathcal{G})^2) \\
&\quad + \beta^i_{t_i-1} \frac{2 \eta_i  + L \eta_i^2}{2} \mathcal{L}^i \mathcal{G}^2 \\
&\quad + \beta^i_{t_i-1} (\tau \eta_i \alpha_t^i \mathcal{L}^i \mathcal{G}^2 + \frac{L}{2} (\tau \eta_i \alpha_t^i \mathcal{L}^i \mathcal{G})^2) \\
&\quad + \mathcal{L}^i (L \eta_i \mathcal{G}^2 + \mathcal{V}^2) \\
= & ~\frac{2 \E\left[ \mathcal{F}(\boldsymbol{w}_{t}) - \mathcal{F}(\boldsymbol{w}_{t+1}) \right]}{\alpha^i_t \eta_i} + \mathcal{L}^i \mathcal{V}^2 \\
&\quad + \frac{\beta^i_{t_i-1} + \tau^2 (\alpha_t^i)^2 \mathcal{L}^i}{2} L \mathcal{L}^i \eta_i^2 \mathcal{G}^2\\
&\quad + (2 \tau \alpha_t^i + L \tau^2  \eta_i (\alpha_t^i)^2 \mathcal{L}^i + \beta^i_{t_i-1} \eta_i \\
&\quad + \beta^i_{t_i-1} \tau + \eta_i \alpha_t^i + L \eta_i) \mathcal{L}^i \mathcal{G}^2.
\end{align*}
We take $\alpha_{min} \leq \alpha_t^i \leq 1$ and $0 \leq \beta^i_{t_i-1} \leq 1$ with $\alpha_{min} > 0$, we can get:
\begin{align*}
&~\sum_{l=1}^{\mathcal{L}^i} \E\parallel \nabla \mathcal{F}(\boldsymbol{w}_{o,l}) \parallel^2 \\
\leq & ~\frac{2 \E\left[ \mathcal{F}(\boldsymbol{w}_{t}) - \mathcal{F}(\boldsymbol{w}_{t+1}) \right]}{\alpha_{min} \eta_i} + \mathcal{L}^i \mathcal{V}^2 + \frac{1 + \tau^2 \mathcal{L}^i}{2} L \mathcal{L}^i \eta_i^2 \mathcal{G}^2\\
& + (3 \tau + L \tau^2  \eta_i \mathcal{L}^i + 2 \eta_i + L \eta_i) \mathcal{L}^i \mathcal{G}^2.
\end{align*}
After $T$ global rounds, we have:
\begin{align*}
&~\frac{1}{\sum_{t=0}^{T}\mathcal{L}_t} \sum_{t=0}^{T} \sum_{l=0}^{\mathcal{L}_t}\E\parallel \nabla \mathcal{F}(\boldsymbol{w}_{o,l}) \parallel^2 \\
\leq &~\frac{2 \E\left[ \mathcal{F}(\boldsymbol{w}_{0}) - \mathcal{F}(\boldsymbol{w}_{T}) \right]}{\alpha_{min} \eta_i T \mathcal{L}_{min}} + \frac{1 + \tau^2 \mathcal{L}_{max}}{2T \mathcal{L}_{min}} L \mathcal{L}_{max} \eta_i \mathcal{G}^2\\
& + \frac{3 \tau + L \tau^2  \eta_i \mathcal{L}_{max} + 2 \eta_i + L \eta_i}{T \mathcal{L}_{min}}\mathcal{L}_{max} \mathcal{G}^2 + \frac{\mathcal{L}_{max} \mathcal{V}^2}{T \mathcal{L}_{min}},
\end{align*}
where $\mathcal{L}_t$ represents the maximum local epochs within the $t$-th global round with $\mathcal{L}_{min} \leq \mathcal{L}_t \leq \mathcal{L}_{max}$. We take  $\eta_i = \frac{1}{\sqrt{T}}$ and $T = \mathcal{L}_{min}^6$, and can get:
\begin{align*}
&~\frac{1}{\sum_{t=0}^{T}\mathcal{L}_t} \sum_{t=0}^{T} \sum_{l=0}^{\mathcal{L}_t}\E\parallel \nabla \mathcal{F}(\boldsymbol{w}_{o,l}) \parallel^2 \\
\leq & ~\frac{2 \E\left[ \mathcal{F}(\boldsymbol{w}_{0}) - \mathcal{F}(\boldsymbol{w}_{T}) \right]}{\alpha_{min} \mathcal{L}_{min}^3} + \frac{1 + \tau^2 \mathcal{L}_{max}}{2 \mathcal{L}_{min}^3} L \mathcal{L}_{max} \mathcal{G}^2\\
&\quad + \frac{3 \tau \mathcal{L}_{max} \mathcal{G}^2 + \mathcal{L}^i \mathcal{V}^2}{ \mathcal{L}_{min}^7} + \frac{ L \tau^2 \mathcal{L}_{max} + 2 + L }{\mathcal{L}_{min}^3}\mathcal{L}_{max} \mathcal{G}^2 \\
\leq & \frac{2 \E\left[ \mathcal{F}(\boldsymbol{w}_{0}) - \mathcal{F}(\boldsymbol{w}_{T}) \right]}{\alpha_{min} \mathcal{L}_{min}^3} + \OM\left( \frac{L \mathcal{G}^2 \mathcal{L}_{max}}{\mathcal{L}_{min}^3} \right) \\
&\quad + \OM\left( \frac{\mathcal{L}^i \mathcal{V}^2}{ \mathcal{L}_{min}^7}\right) + \OM\left( \frac{\tau \mathcal{G}^2 \mathcal{L}_{max}}{ \mathcal{L}_{min}^7} \right) \\
&\quad + \OM\left( \frac{ \mathcal{G}^2 \mathcal{L}_{max} }{\mathcal{L}_{min}^3} \right) + \OM\left( \frac{ L \mathcal{G}^2 \mathcal{L}_{max} }{\mathcal{L}_{min}^3} \right)\\
&  + \OM\left(\frac{ L \tau^2 \mathcal{G}^2 \mathcal{L}_{max}^2}{\mathcal{L}_{min}^3} \right) + \OM\left( \frac{ L \tau^2 \mathcal{G}^2 \mathcal{L}_{max}^2 }{\mathcal{L}_{min}^3} \right) 
\end{align*}

\end{proof}

\subsection{Experiment details}

\begin{table*}[!t]
\centering
\caption{The network structure of CNN.}
\label{tbl:cnn}
\begin{tabular}{|l|l|l|}
\hline 
Layer (type) & Parameters & Input Layer \\ \hline 
conv1(Convolution)& channels=64, kernel\_size=2 &data \\ \hline 
activation1(Activation)& null &conv1 \\ \hline 
conv2(Convolution)& channels=32, kernel\_size=2 &activation1 \\ \hline 
activation2(Activation)& null &conv2 \\ \hline 
flatten1(Flatten)& null &activation2 \\ \hline 
dense1(Dense)& units=10 &flatten1 \\ \hline 
softmax(SoftmaxOutput)& null &dense1 \\ \hline 
\end{tabular} 
\end{table*}

In the experiment, we exploit a CNN model with the network structure shown in Table~\ref{tbl:cnn}. We exploit 44 Tesla V100 GPU cards to simulate the FL environment. We simulate device heterogeneity by considering the variations in local training times, i.e., the training time of the slowest device is five times longer than that of the fastest device, and the training time of each device is independently and randomly sampled within this range. We exploit a learning rate decay for the training process. In addition, we take 500 as the maximum number of epochs for synchronous approaches and 5000 as that of asynchronous approaches. The server triggers one idle task every 5 seconds, with a maximum parallelism constraint, i.e., 10\% of the total device number. We fine-tune the hyper-parameters for each approach and report the best one in the paper.  The summary of main notations is shown in Table \ref{tab:summary} and the values of hyper-parameters are shown in Tables \ref{tab:parameters1} and \ref{tab:parameters2}.

\begin{table*} 
\caption{Summary of main notations.}
\label{tab:summary}
\begin{center}
\begin{tabular}{cc}
\hline
Notation & Definition \\
\hline

$\mathcal{M}$; $m$ & The set of edge devices; the size of $\mathcal{M}$ \\
$\mathcal{D}$; $|\mathcal{D}|$ & The global dataset; the size of $\mathcal{D}$\\
$\mathcal{D}_i$; $|\mathcal{D}_i|$ &  The dataset on Device $i$; the size of $\mathcal{D}_i$\\
$\mathcal{F}(\cdot)$; $\mathcal{F}_i(\cdot)$ & The global loss function; the local loss function on Device $i$ \\
$T$ & The maximum number of global rounds \\
$\mathcal{L}^i$& The maximum number of local epochs on Device $i$ \\
$\tau$& The maximum staleness \\
$\mathcal{T}$ & The constant time period to trigger devices\\
$m'$ & The number of devices to trigger within each time period  \\
$\boldsymbol{w}_t$ & The global model of Version $t$\\
$\boldsymbol{w}_o^i$ & The updated local model from Device $i$ with the original version $o$\\
$\boldsymbol{w}_{o,l}$ & The updated local model with the original version $o$ at local epoch $l$\\
$\boldsymbol{w}_g$ & The fresh global model of Version $g$\\
$\lambda_t^i$, $\sigma_t^i$, $\iota_t^i$ & The control parameters of Device $i$ within the $t$-th local training on the Server\\
$\eta_{\lambda^i}$, $\eta_{\sigma^i}$, $\eta_{\iota^i}$ & The learning rates to update control parameters for Device $i$ on the Server\\
$\alpha^i_t$  & The weight of updated local model from Device $i$ and the $t$-th local training \\
$\beta^i_{t_i}$  & The weight of fresh global model on Device $i$ for the $t_i$-th local model aggregation \\
$\gamma^i_{t_i}$, $\upsilon^i_{t_i}$ & The control parameters of Device $i$ for the $t_i$-th local model aggregation\\
$\eta_{\gamma^i}$, $\eta_{\upsilon^i}$ & The learning rates to update control parameters for Device $i$ on devices\\
$\eta_i$ & The learning rate on Device $i$ \\
$\eta^i_{RL}$ & The learning rate for the update of RL on Device $i$ \\
$\Theta_t$ & The parameters in the RL model at global round $t$\\

\hline
\end{tabular}
\end{center}
\end{table*}


\begin{table*}[htbp]
\caption{Values of hyper-parameters in the experimentation.}
\label{tab:parameters1}
\begin{center}
\begin{tabular}{c|c|c|c|c|c|c|c}
\toprule
\multicolumn{1}{c|}{\multirow{3}{*}{Name}} & \multicolumn{7}{c}{Values} \\
\cline{2-8}
\multicolumn{1}{c|}{}& \multicolumn{3}{c|}{LeNet} & \multicolumn{2}{c|}{CNN} & \multicolumn{2}{c}{ResNet} \\
\cline{2-8}
\multicolumn{1}{c|}{} & FMNIST & CIFAR-10 & CIFAR-100  & CIFAR-10 & CIFAR-100 & CIFAR-100 & Tiny-ImageNet  \\
\hline
$m$ & 100 & 100 & 100 & 100 & 100 & 100 & 100 \\
$m'$ & 10 & 10 & 10 & 10 & 10 & 10 & 10  \\
$T$ & 500 & 500 & 500 & 500 & 500 & 500 & 500 \\
$\tau$ & 99 & 99 & 99 & 99 & 99 & 99 & 99 \\
$\mathcal{T}$ & 10 & 10 & 10 & 10 & 10 & 10 & 10  \\
$\eta_{\lambda^i}$ &0.0001	&0.001	&0.0001	&0.001	&0.00001	&0.0001	&0.0001  \\
$\eta_{\sigma^i}$ &0.0001	&0.001	&0.0001	&0.001	&0.00001	&0.0001	&0.0001 \\
$\eta_{\iota^i}$ &0.0001	&0.1	&0.0001	&0.0001	&0.00001	&0.0001	&0.0001 \\
$\eta_{\gamma^i}$ &0.0001	&0.1	&0.0001	&0.1	&0.00001	&0.0001	&0.0001  \\
$\eta_{\upsilon^i}$&0.0001	&0.001	&0.0001	&0.001	&0.00001	&0.0001	&0.0001   \\
$\eta_i$ &0.005	&0.03	&0.03	&0.028	&0.013	&0.03	&0.03  \\
$\eta^i_{RL}$ & 0.001 & 0.001 & 0.001 & 0.001 & 0.001 & 0.001 & 0.001  \\
\bottomrule
\end{tabular}
\end{center}
\end{table*}

\begin{table*}[htbp]
\caption{Values of hyper-parameters in the experimentation.}
\label{tab:parameters2}
\begin{center}
\begin{tabular}{c|c|c|c|c|c}
\toprule
\multicolumn{1}{c|}{\multirow{3}{*}{Name}} & \multicolumn{5}{c}{Values} \\
\cline{2-6}
\multicolumn{1}{c|}{}& \multicolumn{2}{c|}{AlexNet} & \multicolumn{2}{c|}{VGG} & \multicolumn{1}{c}{TextCNN} \\
\cline{2-6}
\multicolumn{1}{c|}{}& CIFAR-10 & CIFAR-100 & CIFAR-10 & CIFAR-100 & IMDb \\
\hline
$m$ & 100 & 100 & 100 & 100 & 100  \\
$m'$ & 10 & 10 & 10 & 10 & 10  \\
$T$ & 500 & 500 & 500 & 500 & 500 \\
$\tau$ & 99 & 99 & 99 & 99 &99\\
$\mathcal{T}$ & 10 & 10 & 10 & 10 & 10 \\
$\eta_{\lambda^i}$ &0.0001	&0.0001	&0.0001	&0.0001	&0.0001 \\
$\eta_{\sigma^i}$ &0.0001	&0.0001	&0.0001	&0.0001	&0.0001 \\
$\eta_{\iota^i}$ &0.0001	&0.0001	&0.0001	&0.0001	&0.0001 \\
$\eta_{\gamma^i}$  &0.0001	&0.0001	&0.0001	&0.0001	&0.0001                        \\
$\eta_{\upsilon^i}$  &0.0001	&0.0001	&0.0001	&0.0001	&0.0001                      \\
$\eta_i$ &0.03	&0.03	&0.03	&0.03	&0.001 \\
$\eta^i_{RL}$ & 0.001 & 0.001 & 0.001 & 0.001 & 0.001 \\
\bottomrule
\end{tabular}
\end{center}
\end{table*}

\subsection{Visualization of Experimental Results}

\begin{figure*}[!t]
\centering
\subfigure[LeNet \& CIFAR-10]{
\includegraphics[width=0.3\linewidth]{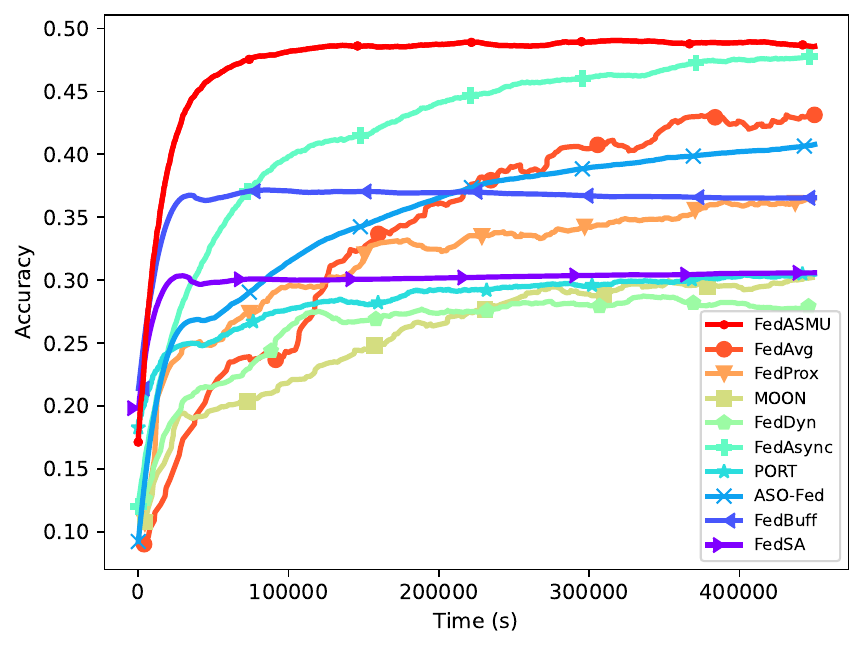}
\label{fig:cmp_dh_lenet_10}
}
\subfigure[CNN \& CIFAR-10]{
\includegraphics[width=0.3\linewidth]{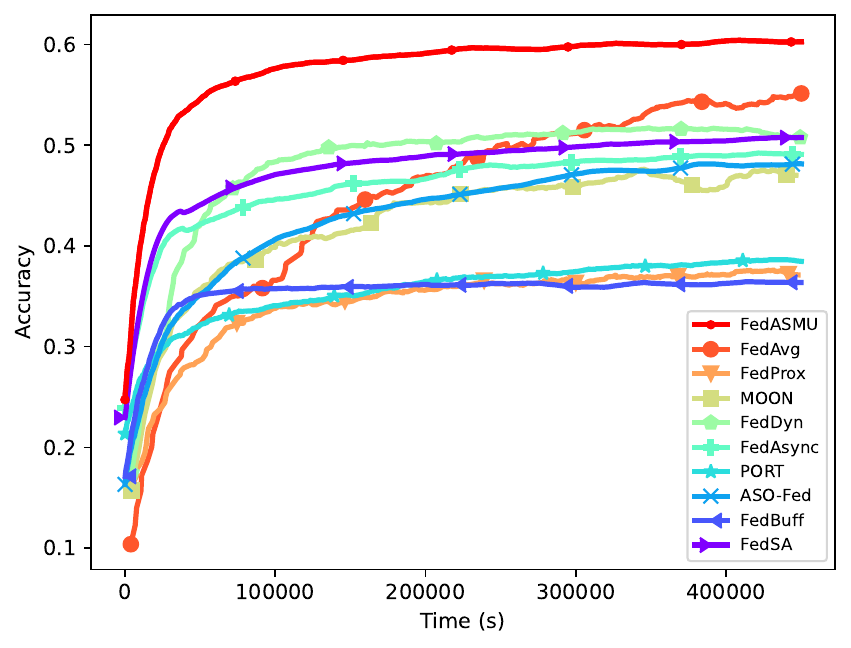}
\label{fig:cmp_dh_cnn_10}
}
\subfigure[ResNet \& Tiny-ImageNet]{
\includegraphics[width=0.3\linewidth]{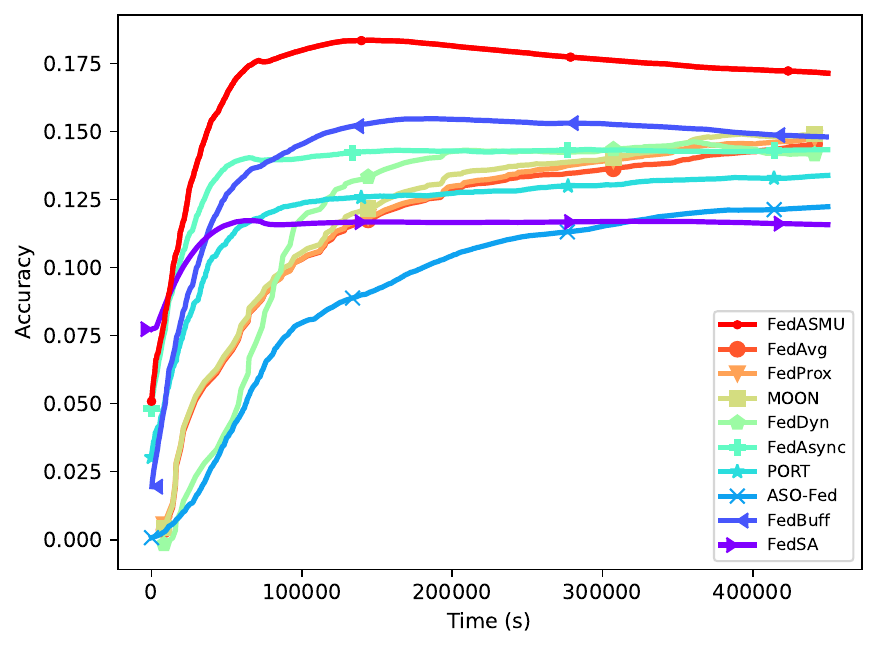}
\label{fig:cmp_dh_resnet_tiny}
}
\caption{The accuracy and training time for \TheName{} and baseline approaches with CIFAR-10 and Tiny-ImageNet.}
\label{fig:async_CNN_CIFAR10}
\end{figure*}

\begin{figure*}[!t]
\centering
\subfigure[LeNet \& CIFAR-100]{
\includegraphics[width=0.3\linewidth]{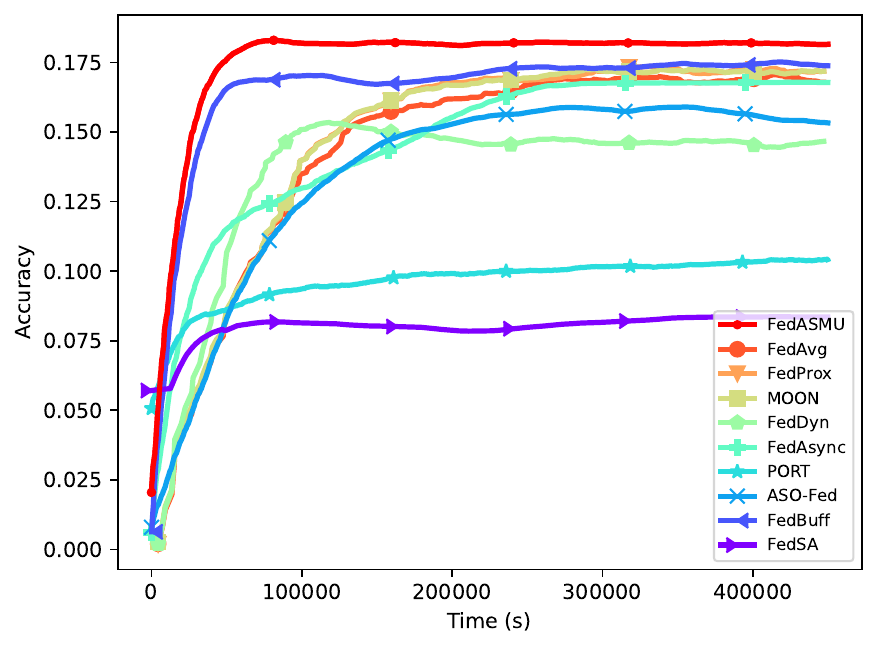}
\label{fig:cmp_dh_lenet_cifar100}
}
\subfigure[CNN \& CIFAR-100]{
\includegraphics[width=0.3\linewidth]{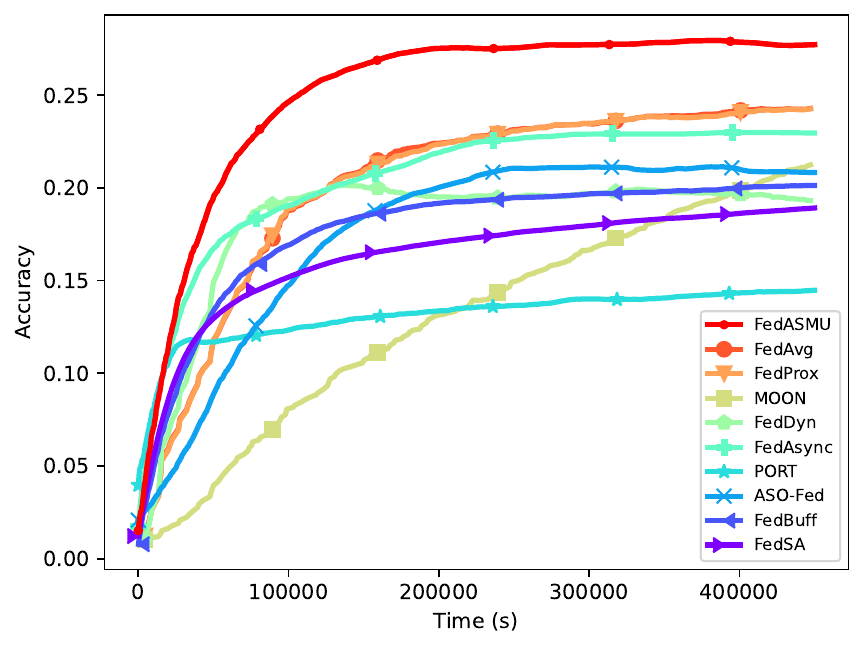}
\label{fig:cmp_dh_cnn_cifar100}
}
\subfigure[ResNet \& CIFAR-100]{
\includegraphics[width=0.3\linewidth]{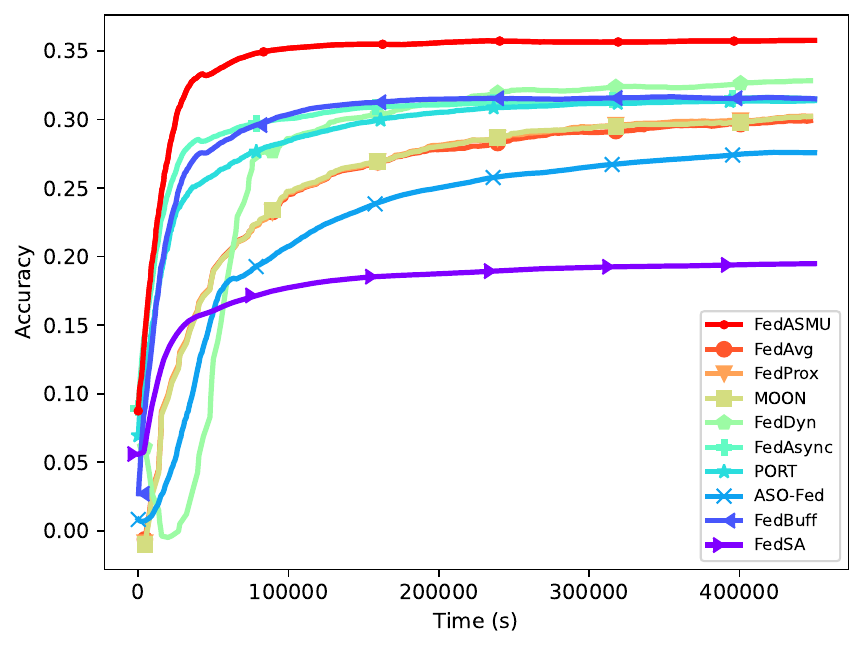}
\label{fig:cmp_dh_resnet_cifar100}
}
\caption{The accuracy and training time for \TheName{} and baseline approaches with CIFAR-100.}
\label{fig:async_cmp}
\end{figure*}

\begin{figure*}[!htbp]
\centering
\subfigure[AlexNet \& CIFAR-10]{
\includegraphics[width=0.3\linewidth]{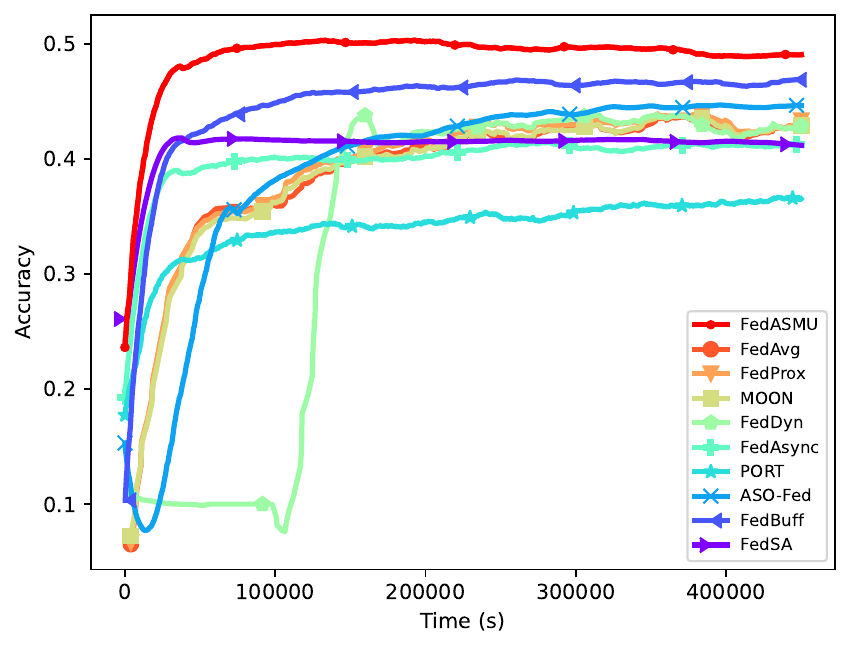}
\label{fig:alexnet_CIFAR10_cmp}
}
\subfigure[VGG \& CIFAR-10]{
\includegraphics[width=0.3\linewidth]{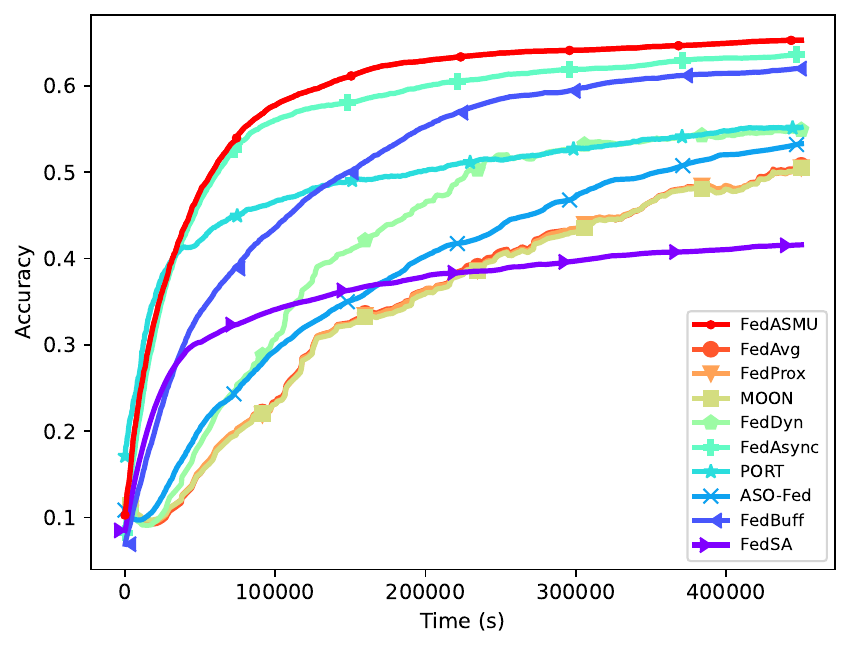}
\label{fig:vgg11_CIFAR10_cmp}
}
\subfigure[TextCNN \& IMDb]{
\includegraphics[width=0.3\linewidth]{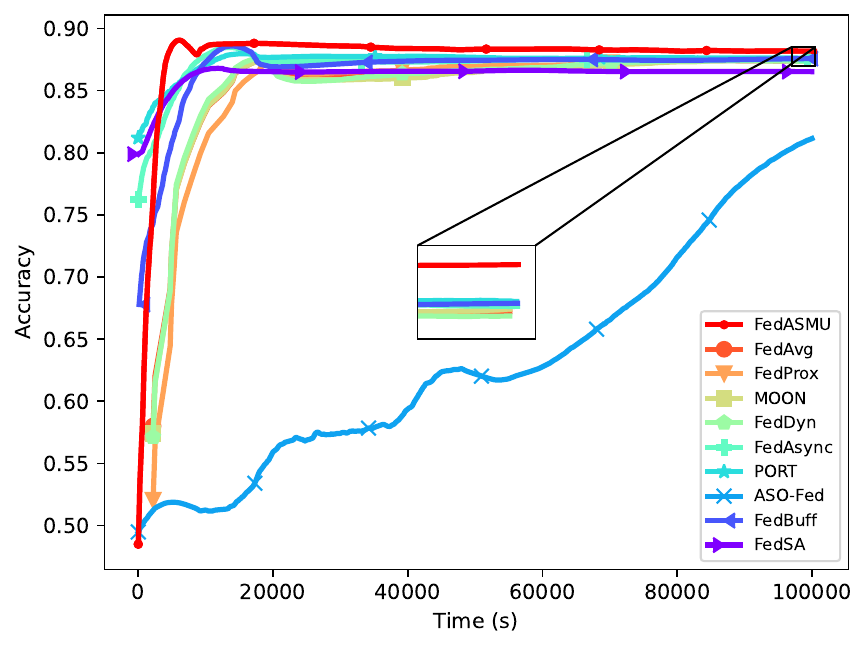}
\label{fig:textcnn_cmp}
}
\caption{The accuracy and training time for \TheName{} and baseline approaches with CIFAR-10 and IMDb}
\label{fig:async_cmp_cifar-10-IMDb}
\end{figure*}

\begin{figure*}[!htbp]
\centering
\subfigure[AlexNet \& CIFAR-100]{
\includegraphics[width=0.3\linewidth]{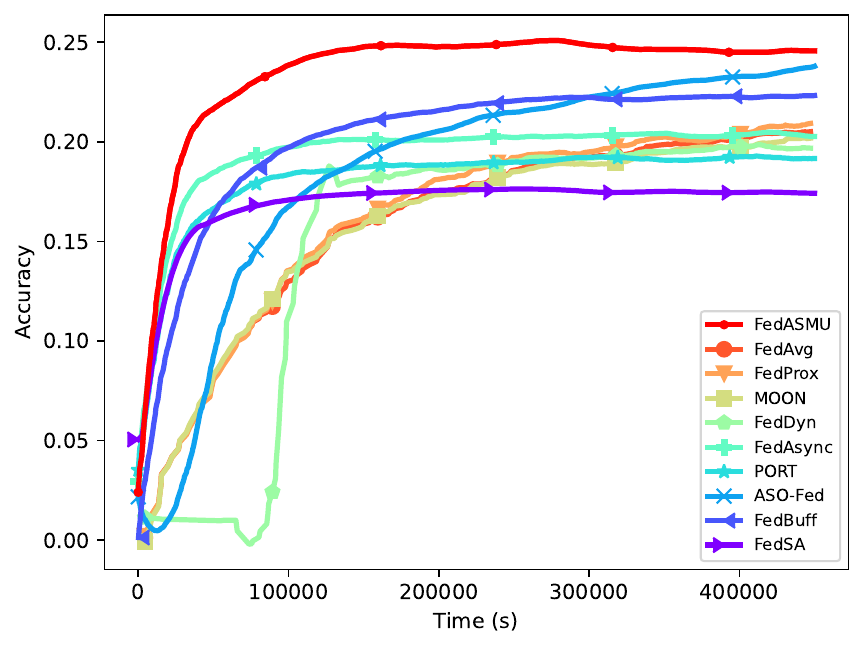}
\label{fig:alexnet_CIFAR100_cmp}
}
\subfigure[VGG \& CIFAR-100]{
\includegraphics[width=0.3\linewidth]{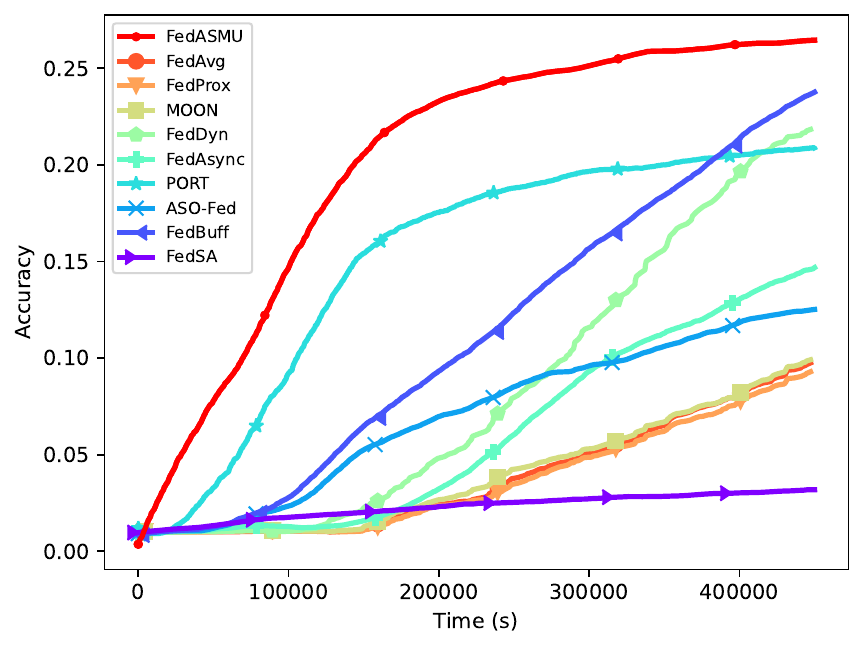}
\label{fig:vgg11_CIFAR100_cmp}
}
\subfigure[LeNet \& FMNIST]{
\includegraphics[width=0.3\linewidth]{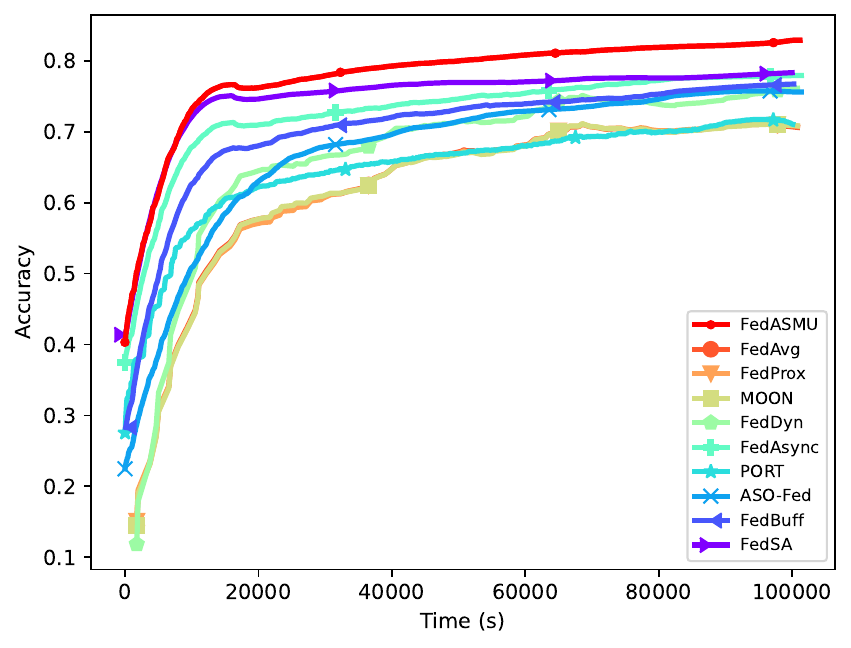}
\label{fig:lenet_FMNIST_cmp}
}
\caption{The accuracy and training time for \TheName{} and baseline approaches with CIFAR-100 and FMNIST.}
\label{fig:async_cmp_cifar-100-FMNIST}
\end{figure*}

\begin{figure*}[!htbp]
\centering
\subfigure[Diversified device hetergeneity]{
\includegraphics[width=0.3\linewidth]{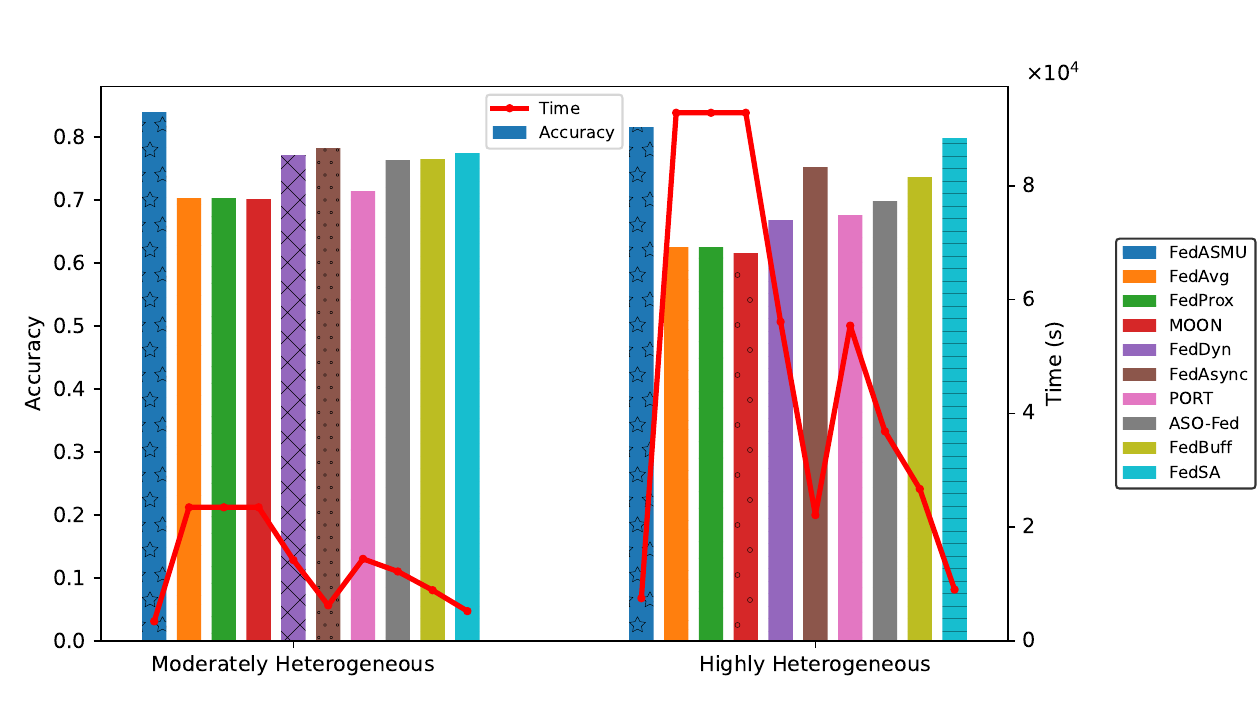}
\label{fig:heter_lenet_FMNIST}
}
\subfigure[Different network bandwidth]{
\includegraphics[width=0.3\linewidth]{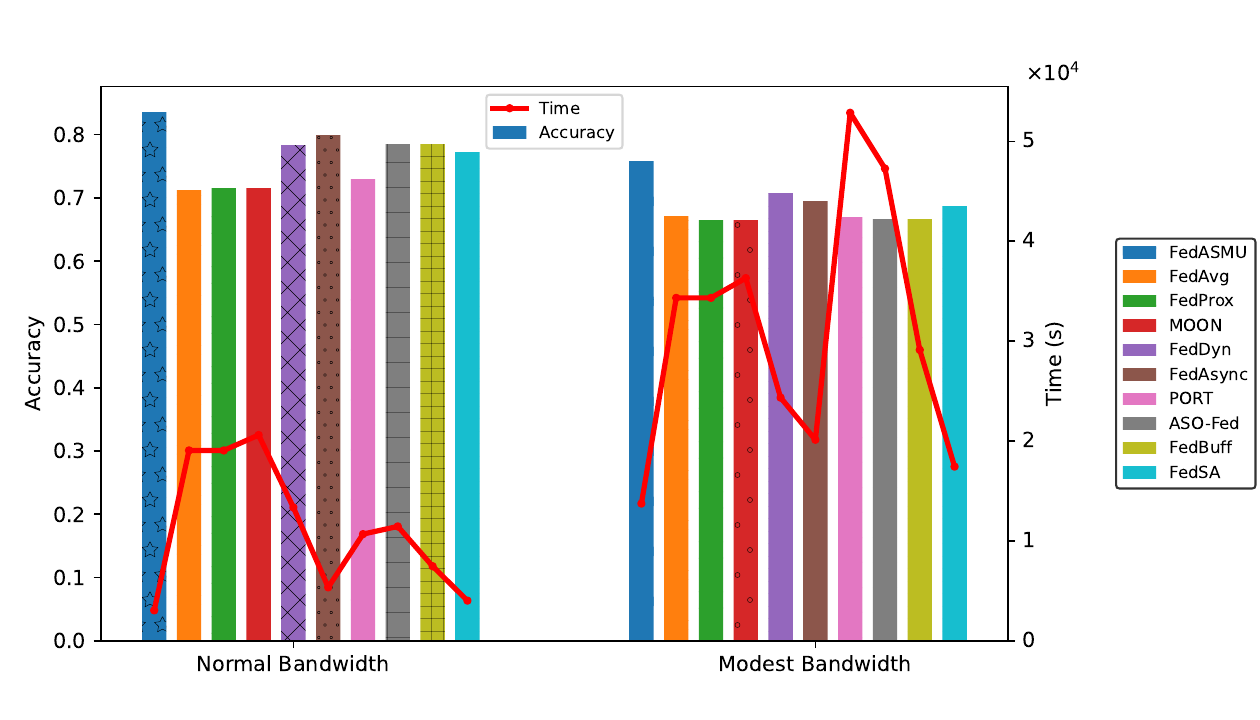}
\label{fig:bandwidth_lenet_FMNIST}
}
\subfigure[Various numbers of devices]{
\includegraphics[width=0.3\linewidth]{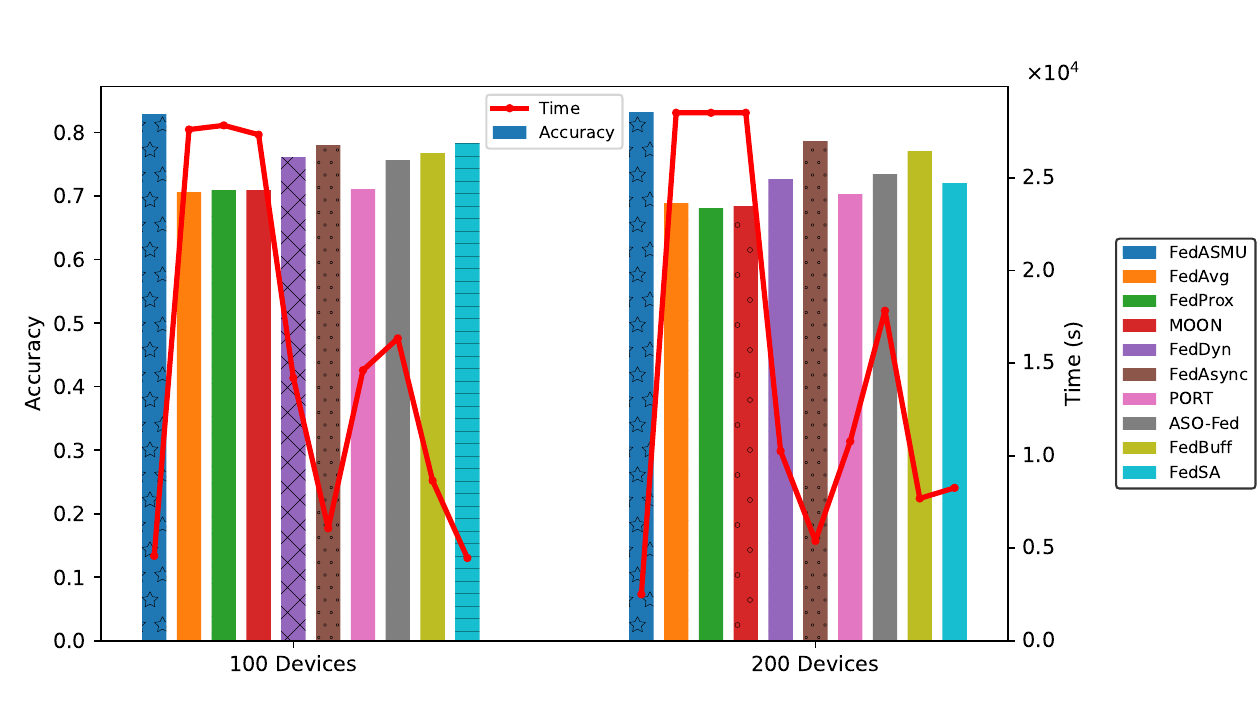}
\label{fig:device_lenet_FMNIST}
}
\caption{The accuracy and training time to achieve target accuracy (0.60) for \TheName{} and baseline approaches with LeNet and FMNIST in diverse environments.}
\label{fig:diverse_environment}
\end{figure*}

The visualization of the experimental results with diverse baseline approaches and normal bandwidth are shown in Figures \ref{fig:async_CNN_CIFAR10}, \ref{fig:async_cmp}, \ref{fig:async_cmp_cifar-10-IMDb}, \ref{fig:async_cmp_cifar-100-FMNIST}. In addition, the visualization of the experimentation within diverse environments, i.e., various numbers of devices, diversified device heterogeneity, and different network bandwidth, are shown in Figure \ref{fig:diverse_environment}. First, as shown in Figure \ref{fig:heter_lenet_FMNIST}, we verify that \TheName{} can still outperform baseline approaches (from 5.04\% to 9.34\% in terms of accuracy and from 21.21\% to 74.01\% in terms of efficiency) when the network becomes modest (50 times lower than the normal network bandwidth). Then, we vary the heterogeneity of devices to show that \TheName{} can well address the heterogeneity with superb accuracy and high efficiency, by augmenting the difference (from 110 times faster to 440 times faster) between the fastest device and the lowest device while randomly sample the local training time for the other devices, as shown in Figure \ref{fig:bandwidth_lenet_FMNIST}. Finally, we carry out experiments with 100 and 200 devices to show that \TheName{} corresponds to excellent scalability as shown in Figure \ref{fig:device_lenet_FMNIST}.

\subsection{Communication Overhead Analysis}

The additional communication overhead of \TheName{} mainly lies in the downloading global models in the down-link channel from server to devices. Since the down-link channel has high bandwidth, which incurs acceptable extra costs with significant benefits (higher accuracy and shorter training time). To analyze the performance of \TheName{}, we carry out extra experimentation with the bandwidth of 100 (100 times smaller than normal), the advantages of \TheName{} becomes even more significant compared with 50 (50 times smaller) (5.04\%-9.34\% for 50 to 1.4\%-\textbf{12.6}\% for 100) in terms of accuracy and (21.21\%-62.17\% for 50 to 6.7\%-\textbf{71.9}\% for 100) in terms of training time, which reveals excellent performance of \TheName{} within modest network environments.

\subsection{Hyper-parameter Fine-tuning}

We conduct extra experiments on FMNIST and LeNet with varying trigger periods $\mathcal{T}$, $\eta_{\lambda^i}$, $\eta_{\sigma^i}$, and $\eta_{\iota^i}$, which correspond to little difference (0.0\% to 0.8\% with only one exceptional case of 2.4\%) thanks to our dynamic model aggregation. Thus, \TheName{} is not sensitive to the hyper-parameters and easy to fine-tune.

\subsection{Comparison with Other Baselines}

We carry out extra experimentation to compare FedASMU with three more recent works in asynchronous FL, i.e., FedDelay \cite{Koloskova2022SharperCG}, SyncDrop \cite{Dun2022EfficientAL}, and AsyncPart \cite{Wang2021AsynchronousFL}. We find FedASMU significantly outperforms these three approaches in terms of accuracy (0.063\% for FedDelay, 11.2\% for SyncDrop, and 6.7\% for AsyncPart) and training time for a target accuracy (69.2\% for FedDelay, 19.4\% for SyncDrop, and 75.9\% for AsyncPart).

\subsection{Ablation Study for Request Time Slot Selection}

We carry out extra experimentation with three heuristics. H1: the device sends the request just after the first local epoch. H2: the device sends the request in the middle of the local trainng. H3: the device sends the request at the last but one local epoch. The accuracy of our RL approach is significantly higher than H1 (2.4\%), H2 (2.2\%), and H3 (2.8\%).

\end{document}